\documentclass[sigconf]{acmart}

\usepackage{microtype}
\usepackage{graphicx}
\usepackage{subfigure}
\usepackage[labelformat=simple]{subcaption}
\usepackage{booktabs} 
\usepackage{multirow}
\usepackage{longtable}

\usepackage{hyperref}
\usepackage{caption}
\usepackage{xcolor}
\usepackage{color}

\usepackage{amsmath,amscd,url,enumerate, mathtools}
\usepackage[utf8]{inputenc}
\usepackage{wrapfig}
\usepackage{pdflscape}
\usepackage{tikz}
\usetikzlibrary{decorations.markings,arrows}
\usetikzlibrary{positioning,arrows,fit}

\usepackage{tikz-cd}
\usepackage{amsfonts}
\usepackage{listings}
\usepackage{graphicx}
\usepackage{mathrsfs}
\usepackage{dsfont}
\usepackage{enumitem}
\usepackage{array}
\usepackage{verbatim}
\newcolumntype{?}{!{\vrule width 1pt}}
\usepackage[capitalize,noabbrev]{cleveref}
\usepackage[font=footnotesize,labelfont=bf, labelsep=period, textfont=sl]{caption}
\usepackage{pythonhighlight} 

\definecolor{copper}{HTML}{C84E00}
\definecolor{dandelion}{HTML}{FFD960}
\definecolor{piedmont}{HTML}{A1B70D}
\definecolor{eno}{HTML}{339898}
\definecolor{shale}{HTML}{0577B1}
\definecolor{dukeblue}{HTML}{012169}
\definecolor{ironweed}{HTML}{993399}
\definecolor{whisper}{HTML}{F3F2F1}
\definecolor{gingerbeer}{HTML}{FCF7E5}

\hypersetup{
  colorlinks=true,
  linkcolor=dukeblue!70!white,
  urlcolor=dukeblue!70!white,
  citecolor=dukeblue!70!white
}

\usepackage{tcolorbox}
\tcbuselibrary{most}

\newtcolorbox{callout}{
    colback=gingerbeer!80,
    colframe=black,
    coltext=black,
    boxrule=1pt,
    arc=5pt,
    left=5pt,
    right=5pt,
    top=5pt,
    bottom=5pt,
}

\newcommand{\para}[1]{{\vspace{2pt} \bf \noindent #1}}  
\newenvironment{packed_itemize}{
\begin{list}{\labelitemi}{\leftmargin=1em}
\setlength{\itemsep}{1pt}                                                           
\setlength{\parskip}{0pt}                                                                                 \setlength{\parsep}{0pt}                                                                                  \setlength{\headsep}{0pt}                                                                                 \setlength{\topskip}{0pt}                                                                                 \setlength{\topmargin}{0pt}                                                                               \setlength{\topsep}{0pt}                                                                                  \setlength{\partopsep}{0pt}                                                                               }{\end{list}}

\newcommand{\rbts}{\texttt{robots.txt}~}

\copyrightyear{2025}
\acmYear{2025}
\setcopyright{cc}
\setcctype{by}
\acmConference[IMC '25]{Proceedings of the 2025 ACM Internet Measurement Conference}{October 28--31, 2025}{Madison, WI, USA}
\acmBooktitle{Proceedings of the 2025 ACM Internet Measurement Conference (IMC '25), October 28--31, 2025, Madison, WI, USA}
\acmDOI{10.1145/3730567.3764471}
\acmISBN{979-8-4007-1860-1/2025/10}

\begin{document}

\title[]{Scrapers Selectively Respect \texttt{robots.txt} Directives:\\ Evidence From a Large-Scale Empirical Study}

\author{Taein Kim}
 \affiliation{%
 \department{Department of Electrical and Computer Engineering}
  \institution{Duke University}
  \city{Durham}
  \state{NC}
  \country{USA}}

\author{Karstan Bock}
 \affiliation{%
 \department{Department of Electrical and Computer Engineering}
  \institution{Duke University}
  \city{Durham}
  \state{NC}
  \country{USA}}

\author{Claire Luo}
 \affiliation{%
 \department{Department of Electrical and Computer Engineering}
  \institution{Duke University}
  \city{Durham}
  \state{NC}
  \country{USA}}

\author{Amanda Liswood}
 \affiliation{%
 \department{Department of Electrical and Computer Engineering}
  \institution{Duke University}
  \city{Durham}
  \state{NC}
  \country{USA}}

  \author{Chloe Poroslay}
 \affiliation{%
 \department{Office of Information Technology}
  \institution{Duke University}
  \city{Durham}
  \state{NC}
  \country{USA}}

  \author{Emily Wenger}
\authornote{Corresponding author: emily.wenger@duke.edu}
 \affiliation{%
 \department{Department of Electrical and Computer Engineering}
  \institution{Duke University}
  \city{Durham}
  \state{NC}
  \country{USA}}

\renewcommand{\shortauthors}{Taein Kim et al.}

\begin{abstract}
 Online data scraping has taken on new dimensions in recent years, as traditional scrapers have been joined by new AI-specific bots. To counteract unwanted scraping, many sites use tools like the Robots Exclusion Protocol (REP), which places a \rbts file at the site root to dictate scraper behavior. Yet, the efficacy of the REP is not well-understood. Anecdotal evidence suggests some bots comply poorly with it, but no rigorous study exists to support (or refute) this claim. To understand the merits and limits of the REP, we conduct the first large-scale study of web scraper compliance with \rbts directives using anonymized web logs from our institution. We analyze the behavior of $130$ self-declared bots (and many anonymous ones) over $40$ days, using a series of controlled \rbts experiments. We find that bots are less likely to comply with stricter \rbts directives, and that certain categories of bots, including AI search crawlers, rarely check \rbts at all. Our findings suggest that relying on \rbts to prevent unwanted scraping is risky and highlight the need for alternatives.
\end{abstract}



\maketitle

\section{Introduction}

Web scraping, the process of systematically extracting and downloading information from websites, has been a part of the internet ecosystem since its early days~\cite{wanderer}. Scraping is now a critical part of many companies' business models, allowing search engines to optimize results and shopping sites to compare deals, among other purposes. In recent years, scraping has gained new importance as vast troves of internet data are tapped for a novel purpose: training and operating large-scale AI models. Today's large-scale AI models require  terabytes of training data~\cite{kaplan2020scaling, oai_rfi_2025}, for which the internet is an obvious and cheap source. Whitepapers documenting many of today's biggest models, from GPT to Llama and beyond, openly acknowledge the use of web scraping to create training data~\cite{team2024jamba, brown2020language, dubey2024llama, touvron2023llama, abdin2024phi, chowdhery2023palm}. 

The AI demand for web scraping extends beyond training data. AI models for text generation have a pesky tendency to {\em hallucinate}, confidently stating incorrect information~\cite{ji2023survey, raunak2021curious}. To make AI-generated text  more reliable, AI developers proposed Retrieval-Augmented Generation (RAG), in which large vectorized databases of web content are used to ground AI responses~\cite{guu2020retrieval, lewis2020retrieval, izacard2020leveraging}. These databases are collated via large-scale web scraping and, unlike training datasets, are continuously updated. Additionally, AI ``agents,'' generative models with additional capabilities, can deploy bots to fetch web content as part of their workflow~\cite{claude_internet, gemini_internet, gpt_interent}. 

Beyond well-documented copyright and privacy concerns, widespread scraping for any purpose\textemdash including for AI training data, RAG, and agent use\textemdash can destabilize public, data-rich websites. Numerous writings from the ``grey'' (non-academic) literature describe sites being taken offline in recent months due to thousands of download requests from scrapers associated with AI companies~\cite{perplexity_wired, amazon-bots, anthropic_bots}. These scraping-induced problems have prompted an escalating tactical war, as content-rich websites attempt to moderate or prevent unwanted web scraping~\cite{tarpits, longpre2024consent, crawler_wars}. 

Among the most widely used anti-scraping solution is the Robots Exclusion Protocol (REP), implemented via a \rbts file located at the site root. The REP lets site owners set rules that specify which bots can scrape information from the site, which sub-domains those bots can access, and how long bots must wait between successive page accesses~\cite{rep}.  Recent work~\cite{longpre2024consent} showed a significant uptick in the use of and restrictions in \rbts files after the rise of generative AI models around 2022, purportedly in reaction to increased scraping for training data. The REP was codified in 2022, but it is not legally binding, so compliance requires scraper goodwill.

Although \rbts is widely used, little hard evidence exists to prove its efficacy in preventing scraping. Some gray literature suggests that many crawlers, including AI-specific bots, do not respect \rbts~\cite{anthropic_bots}. Other articles claim that certain crawlers ignore \rbts files altogether~\cite{perplexity_wired, tarpits}, while still others observe that bots disallowed by \rbts sometimes pretend to be other, sanctioned bots to circumvent restrictions~\cite{amazon-bots}. Despite this, a recent work~\cite{liu2024somesite} conducted a study of $7$ crawlers and observed that all AI bots that promised to respect \rbts did so.  

 Given the dearth of options for deterring unwanted web scraping and widespread belief that \rbts can help, there is urgent need for a large-scale, controlled experiment evaluating this belief. Such a study will allow content hosts to make informed decisions about whether \rbts will provide their desired level of protection or if alternative deterrence mechanisms are needed.

 \para{Our contribution.} We conduct the first controlled study of scraper compliance with \rbts at scale. To do this, we collect anonymized web traffic data from a set of 36 websites we control over $40$ days. During this period, we conduct two empirical measurement studies on sites' \rbts files to determine if or to what extent web bots, AI and otherwise, comply with \rbts directives. 

The first study, conducted on a single site in our dataset with significant web traffic, deployed three versions of a \rbts file with increasingly strict bot directives, each for two weeks. By analyzing how bot behavior changes in response to different \rbts directives, we can determine which directives, if any, effectively deter bot behavior and which bots are more or less respectful of \rbts. The second study, conducted via passive observation of bot behavior on sites with fixed \rbts files, broadens the perspective on the analysis, determining the frequency with which \rbts files are accessed. Along the way, we make observations about bot and scraper behavior in general that we believe the community will find interesting.  Our key findings are as follows:
\begin{packed_itemize}
    \item {\bf Bots are less likely to respect \rbts that employ strict directives,} such as denying access to certain pages. Across our three versions of \rbts, we see compliance decrease as directives become stricter. 
    \item {\bf SEO bots are most respectful of \rbts, while search engine crawlers are among the least.} AI-specific bots like AI assistants and AI data scrapers, fall in between. 
    \item {\bf Observed non-compliance with \rbts among otherwise respectful bots can sometimes be attributed to spoofing,} in which malicious bots present a false user agent to avoid detection. 
\end{packed_itemize}

The rest of the paper is organized as follows. \S\ref{sec:related} situates our work in the broader landscape of data scraping. \S\ref{sec:dataset} describes our data collection process and gives a dataset overview. \S\ref{sec:active} outlines our \rbts experiments and results. \S\ref{sec:mediating} discusses possible confounding variables in our experiments. \S\ref{sec:discussion} lists limitations and future work. 
\section{Context: Web Scraping and Bot Deterrence}
\label{sec:related}

Prior work has studied web scraping in general, as well as the prevalence and properties of \rbts files. In this section, we position our work in the broader landscape of research on data scraping and scraper deterrence.

\subsection{Web scraping: a history}

\para{Early days.} Web scrapers and crawlers are nearly as old as the internet. In this work, crawling refers to the process of systematically accessing links on the web, while scraping is the act of downloading information from each link. The first known web crawler was the World Wide Web Wanderer, created in 1993 by an MIT undergraduate student to measure web growth~\cite{wanderer}. Shortly after this, the first crawler-based search engine, JumpStation, premiered, paving the way for future crawler-fueled search engines like Google~\cite{jumpstation}. The creation of tools such as APIs has allowed for simultaneous scraping and crawling, since APIs enable easy data downloads~\cite{api}. Web crawling and scraping now power a number of useful applications, such as search engines, price comparison, and real-time system monitoring.

\para{Web scraping in the AI era.} Interest in data scraping has only increased as AI models have grown in size and scope. The operating principle for many AI model trainers is that more training data produces better models~\cite{kaplan2020scaling}, and the internet is a free and expansive source of data. To collect data at scale, model trainers can either scrape the data themselves or rely on pre-collected datasets like Common Crawl~\cite{commoncrawl}. Additionally, there are some datasets, like LAION 5B, which only provide URLs, requiring prospective users to scrape the data from the links~\cite{schuhmann2022laion}. Since this practice has become so common, large AI companies have developed new user agents to identify bots scraping data on their behalf~\cite{darkvisitors}.

In recent years, other types of AI-related scraping have emerged: retrieval-augmented generation (RAG) and AI web accesses. RAG~\cite{lewis2020retrieval, guu2020retrieval,izacard2020leveraging} allows large language models (LLMs) to supplement their outputs with current online information, since their training datasets struggle to produce factual information on recent events and can ``hallucinate'' information in their responses~\cite{ji2023survey}. RAG relies on a database of continuously updated web links that models can quickly search. Recent innovations enable generative AI ``agents'' to access websites while completing a user-requested task~\cite{gemini_internet, claude_internet, gpt_interent}. Such web accesses\textemdash which range from information lookup to inputting or downloading data\textemdash typically involve deputizing a scraper bot to perform the task.

\subsection{Bot deterrence}
\label{subsec:deter}

Although certain types of web scraping can benefit site owners (e.g. search engine optimization), other types can be harmful. For example, unfettered scraping can scoop up data that, while publicly available, may have copyright or privacy concerns~\cite{longpre2024consent}. Furthermore, numerous blog posts and other articles from recent years have documented how high-intensity scraping by presumed AI crawlers have caused site instability~\cite{perplexity_wired, amazon-bots, tarpits}. In light of these downsides, many techniques have been proposed to prevent unwanted scraping. 

\begin{table*}[t]
    \vspace{0.2cm}
    \centering
    \begin{tabular}{cc}
    \toprule
        {\bf \rbts field} & {\bf Description} \\ \midrule
        \texttt{user-agent} &  Refers to a bot with self-declared user agent string (e.g. Googlebot).\\ \midrule
        \texttt{allow} & Subset of site pages that bot with specific \texttt{user-agent} {\bf can} access. \\ \midrule
        \texttt{disallow} & Subset of site pages bot with specific \texttt{user-agent} {\bf cannot} access. \\ \midrule
        \texttt{crawl-delay} & Minimum required time between successive page accesses by specific bot. \\ \midrule
        \texttt{sitemap} & Provides an overview of subdomains on the site. \\ \bottomrule
    \end{tabular}
    \vspace{0.1cm}
    \caption{{\bf Common fields in \rbts files.} \rbts files constructed using these fields specify how bots can interact with a site, see Figure~\ref{fig:rbts_ex}.}
    \label{tab:robots_info}
\end{table*}

\begin{figure}
\vspace{-0.3cm}
\rule[1ex]{8.5cm}{0.5pt}
\begin{verbatim}
User-agent: Googlebot
Allow: /
Crawl-delay: 15

User-agent: *
Allow: /allowed-data/
Disallow: /restricted-data/
Crawl-delay: 30

Sitemap: https://X.X.X/sitemap/sitemap-0.xml
\end{verbatim}
\rule[1ex]{8.5cm}{0.5pt}
\vspace{-0.7cm}
\caption{{\bf Example \rbts file}. This site allows bots with \texttt{user-agent} Googlebot to access all subdomains with a \texttt{crawl-delay} of 15 seconds. All other bots are given a \texttt{crawl-delay} of 30 seconds and can only access data under the \texttt{/allowed-data/} subdomain.}
\label{fig:rbts_ex}
\end{figure}

\para{The Robots Exclusion Protocol} emerged in 1994 as a possible solution to unwanted bot activity. As crawlers became prominent on the early web, it became clear that a mechanism was needed to allow ``friendly'' scrapers and disallow poorly-behaved ones~\cite{koster_robots}. The Robots Exclusion Protocol (REP)~\cite{rep} was proposed to fill this gap. The REP asks developers to place a file called \rbts at their site root, specifying restrictions on certain bots and bot behaviors on their site. Possible bot directives include restricting subdomain access, enforcing a crawl delay, and allowing only bots with certain user-agent strings\textemdash see Table~\ref{tab:robots_info} and Figure~\ref{fig:rbts_ex}.  The REP was codified as RFC 9309 by the Internet Engineering Task Force (IETF)~\cite{ietf_rbts} and is widely used~\cite{sun2007large, longpre2024consent}.

Despite its prominence, the REP has significant drawbacks. First, \rbts is not a legally binding document, so requires the good-will of the parties involved for compliance~\cite{rbts_legal}. Second,  it requires web hosts to maintain extensive knowledge of user agents and entities that they wish to exclude from scraping~\cite{longpre2024consent}. This places a heavy burden on hosts to maintain awareness of the ever-changing web landscape. Third, since compliance is optional, some crawlers simply ignore \rbts. Finally, limited work~\cite{liu2024somesite} has studied bot compliance with \rbts, and none have done so at scale, making its efficacy unknown.

\para{Other bot-blocking methods.}  Other methods to block unwanted scraping behavior vary from mildly intrusive to very disruptive. {\em CAPTCHAs}, which ask site visitors to solve a hard (e.g. computationally difficult) problem before viewing or retrieving page content, have historically been effective in deterring bots~\cite{von2003captcha, von2008recaptcha}. However, they add significant friction to user interactions with webpages~\cite{bursztein2010good},  while recent advances in AI (and human skill) make CAPTCHA solving easy~\cite{bursztein2014end, guerar2021gotta, ye2018yet, amin2020web}, both of which make CAPTCHAs less useful. Companies like Cloudflare market {\em proprietary bot detection and deterrence} methods, which can be purchased~\cite{cloudflare_bots}. As a last resort, companies can outright {\em block the IP addresses} of troublesome bots. However, since many bots run on VPNs, IP addresses can be easily recycled, circumventing this approach. Several academic works suggest using {\em reverse proxies} to better regulate traffic to a site~\cite{liu2024somesite, dai2024c}. Finally, other articles suggest novel methods like employing a {\em proof of work}~\cite{amazon-bots} or even a {\em tarpit}~\cite{tarpits}, which creates unending fake content for scrapers.  


\subsection{Prior work on scraping and \rbts}

Numerous studies have considered behaviors of web scrapers~\cite{pham2016understanding, lee2009classification}, but limited work has studied the relationship between scrapers and \rbts.~\citet{longpre2024consent} performed a longitudinal analysis of \rbts files on websites often scraped for AI data and found that  \rbts restrictions tightened after the rise of generative AI models like ChatGPT. Restrictions in these \rbts files vary among AI bots, with large, well-known companies like OpenAI having the most restrictions. Longpre et al. argue that tight controls in robots.txt may degrade the quality of open-source AI training datasets. Concurrent work from~\cite{dinzinger2024longitudinal} similarly observed tightening in \rbts restrictions since the advent of generative AI tools. 

\begin{table*}[t]
\centering
\begin{tabular}{@{}cccccccc@{}}
\toprule
\multicolumn{1}{c}{\bf Data subset} &
  \multicolumn{1}{c}{\begin{tabular}[c]{@{}c@{}}\bf Unique\\ {\bf IP addresses}\end{tabular}} &
  \multicolumn{1}{c}{\begin{tabular}[c]{@{}c@{}}\bf Unique\\ {\bf user agents}\end{tabular}} &
  \multicolumn{1}{c}{\begin{tabular}[c]{@{}c@{}}\bf Avg. bytes \\ {\bf scraped per session}\end{tabular}} &
  \multicolumn{1}{c}{\begin{tabular}[c]{@{}c@{}}\bf Unique\\ {\bf ASNs}\end{tabular}} &
  \multicolumn{1}{c}{\begin{tabular}[c]{@{}c@{}}\bf Total bytes\\ {\bf scraped}\end{tabular}} &
  \multicolumn{1}{c}{\begin{tabular}[c]{@{}c@{}}\bf Total\\ {\bf page visits}\end{tabular}} &
  \multicolumn{1}{c}{\begin{tabular}[c]{@{}c@{}}\bf Unique\\ {\bf page visits}\end{tabular}} \\
\midrule
All data & 231{,}859 & 19{,}250 & 82{,}306 & 8{,}841 & 62{,}713{,}813{,}720 & 761{,}956 & 31{,}665 \\
Known bots & 11{,}291 & 405 & 52{,}612 & 179 & 16{,}706{,}054{,}178 & 317{,}532 & 6{,}347 \\
\bottomrule
\end{tabular}
\caption{{\bf Overview of our dataset}. The top row corresponds to the entire dataset, while the bottom row documents activity associated with well-recognized bots.} 
\label{tab:sites_bots}
\vspace{-0.2cm}
\end{table*}

\para{Do AI bots respect \rbts?} Limited work has focused on whether AI scrapers specifically respect \rbts. One small-scale study explored whether a \rbts file was respected by AI bots and found that, of the $7$ AI bots that visited their site, $6$ promised to respect \rbts and did so~\cite{liu2024somesite}. One (Bytespider from Bytedance) did not promise to respect \rbts and did not. While interesting, the scale of this study prevents it from providing a holistic picture of how bots in general interact with \rbts.

Evidence from grey literature further suggests that noncompliance with \rbts is widespread~\cite{perplexity_wired}. Again, however, no large-scale empirical study has considered this question. We hypothesize this literature gap is primarily due to data access limitations, since to make statistically meaningful statements about \rbts (non)compliance, one must have access to millions of web logs.

\section{Our dataset for analyzing \rbts compliance at scale}
\label{sec:dataset}


\subsection{Dataset Preparation}
\label{sec:data_prep}

To study scraper compliance with \rbts files at scale, we collect and analyze a large dataset composed of network access data from website endpoints managed by our institution, a large private US university. The sites host information about a variety of university functions, from the IT department to campus dining to a personnel directory and beyond. The dataset contains approximately 3.9 million external web requests made to a set of 36 websites from February 12 - March 29, 2025 and includes the following fields:


\begin{packed_itemize}
    \item {\bf Useragent}: A self-reported string identifying the entity accessing a website that often includes the browser, operating system, hardware, and version information of the origin of the request. Well-known bots often provide a simpler identifying string in this field (e.g. Googlebot). 
    \item {\bf Timestamp}: The ISO-8601 formatted time of the request. 
    \item {\bf IP Hash}: A one-way cryptographic hash of the web visitor's IP address. This anonymizes IPs for IRB compliance. 
    \item {\bf Autonomous System Number (ASN)}: A number assigned by the American Registry of Internet Numbers (ARIN) to the entity controlling web visitor's IP. 
    \item {\bf Sitename}: The base website accessed. 
    \item {\bf URI path}: The requested resource. The sitename and URI path combined form the whole URL.
    \item {\bf Status code}: Code site returned in response to request. 
    \item {\bf Bytes}: Amount of data, in bytes, transmitted by the web server during site interaction (e.g. amount of downloaded content). 
    \item {\bf Referer}: Site from which web visitor was redirected (if applicable.)
\end{packed_itemize}
\noindent Each row of the dataset corresponds to one page access by a web visitor at a given time.

\begin{table}[t]
\centering
\begin{tabular}{llll}
\toprule
\textbf{Bot name} & \begin{tabular}[c]{@{}l@{}}{\bf Total} \\\textbf{hits}\end{tabular} & \begin{tabular}[c]{@{}l@{}}\textbf{\% of all}\\{\bf traffic}\end{tabular}  & \begin{tabular}[c]{@{}l@{}}\textbf{GB of data}\\{\bf scraped}\end{tabular} \\ \midrule
YisouSpider         & 121495     & 15.95 & 8.23 \\
Applebot            & 118258     & 15.52 & 0.21 \\
Baiduspider         & 15132      & 1.99  & 0.05 \\
bingbot             & 12900      & 1.69  & 0.80 \\
meta-externalagent  & 12837      & 1.68  & 0.87 \\
Googlebot           & 9103       & 1.19  & 0.85 \\
HeadlessChrome      & 8365       & 1.1   & 1.22 \\
ChatGPT-User        & 3029       & 0.4   & 0.98 \\
yandex.com/bots     & 2179       & 0.29  & 0.28 \\
SemrushBot          & 2111       & 0.28  & 0.06 \\
GPTBot              & 1225       & 0.16  & 0.25 \\
dotbot              & 1066       & 0.14  & 0.01 \\
Amazonbot           & 1009       & 0.13  & 0.07 \\
AhrefsBot           & 862        & 0.11  & 0.02 \\
SkypeUriPreview     & 831        & 0.11  & 0.09 \\
facebookexternalhit & 785        & 0.1   & 0.05 \\
BrightEdge Crawler  & 736        & 0.1   & 0.06 \\
Scrapy              & 726        & 0.1   & 0.19 \\
ClaudeBot           & 684        & 0.09  & 0.09 \\
Bytespider          & 561        & 0.07  & 0.08 \\ \bottomrule
\end{tabular}

\caption{{\bf Over $40\%$ of web accesses in our dataset are attributable to just $20$ bots.} We document the number of unique web accesses for each bot, the proportion of web traffic they compose, and the amount of data each scraped during the 40 day period.}
\label{tab:top_botnames}
\vspace{-0.5cm}
\end{table}

\begin{figure*}[t]
        \begin{minipage}{0.43\linewidth}
        \centering
        \includegraphics[width=\linewidth]{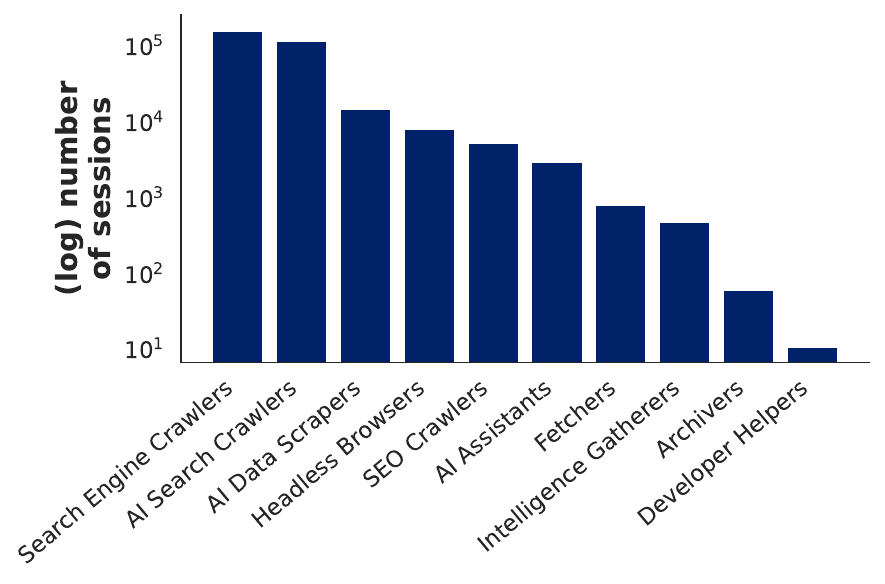}
        \vspace{-0.8cm}
        \caption{{\bf Traditional and AI search, as well as AI data scrapers, are the most active bot types in our dataset}. Headless browsers\textemdash a browser running sans GUI, commonly used by scrapers\textemdash are fourth.}
        \label{fig:hist_bots}
    \end{minipage}
        \hfill
    \begin{minipage}{0.53\linewidth}
        \centering
        \includegraphics[width=\linewidth]{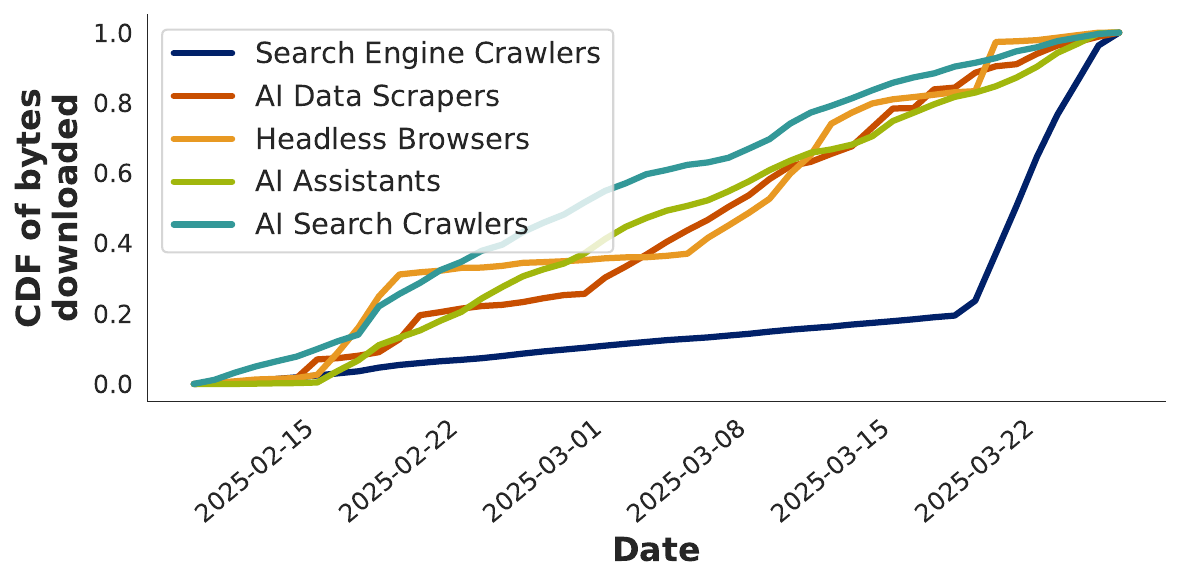}
        \caption{{\bf Most bots in the top $5$ categories in terms of data scraped collect data steadily, but search engine crawlers buck the trend, driven by YisouSpider's mid-March activity}. AI assistants scrape much data relative to session count (Fig~\ref{fig:hist_bots}).}
        \label{fig:total_data}
    \end{minipage}    
    \begin{minipage}{\linewidth}
        \centering
        \vspace{0.2cm}
        \includegraphics[width=\linewidth]{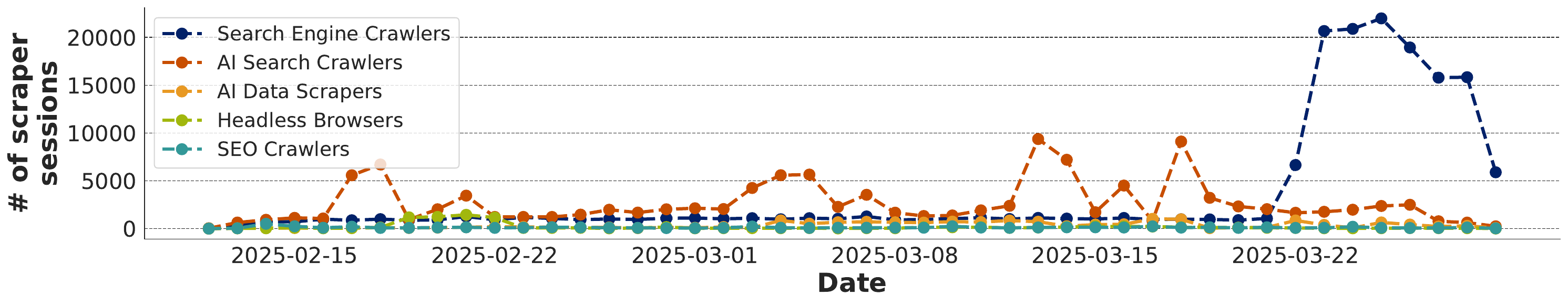}
        \vspace{-0.5cm}
        \caption{{\bf Traditional and AI search crawlers exhibit the most volatility in scraping patterns during the date range encompassed by our dataset.} The volatility in these two categories corresponds directly with high scraping activity of YisouSpider (a search engine crawler) and AppleBot (an AI search crawler). We plot the behaviors of the top $5$ categories of bots by session count for simplicity. }
    \label{fig:scraping_traffic_patterns}
    \end{minipage}
\end{figure*}

\para{Data preprocessing.} Data from any institution-hosted networks or logged-in institutional users is filtered out before it reaches our servers to protect user privacy. We inspect data after collection and remove several IP hashes associated with vulnerability scanning tools and similar entities that are not relevant to our analysis.  We screen out $3$ IP hashes associated with $294,362$ web accesses through this approach. Then, we {\em map ASNs to ARIN info} and {\em standardize bot names} to enhance the data and simplify analysis. 

To map ASNs to ARIN info, we leverage the external library whoisit\footnote{https://pypi.org/project/python-whois/} to poll for whois information for all unique ASNs in our incoming dataset. The returned whois information contains information held by ARIN about the registered user behind the ASN, including the declared entity name and description. We also standardize bot names via fuzzy string matching with a public dataset of common useragent strings and the corresponding bot names\footnote{https://github.com/monperrus/crawler-user-agents}. Using these standardized names, we add an additional column to our dataset that maps bots to bot categories proposed by Dark Visitors, an industry website that keeps an up-to-date list of known scrapers\footnote{https://darkvisitors.com/}\textemdash AI Agents (bots from AI companies with ``agent'' in their name, presumed to operate as part of an agent pipeline), AI Assistants (bots that retrieve content to supplement AI queries), AI Data Scrapers (bots that scrape AI training data), Archivers, Developer Helpers, Fetchers, Headless Agents, Intelligence Gatherers (data collection for non-SEO or AI purposes), Scrapers, Search Engine Crawlers, SEO Crawlers, Uncategorized, and Undocumented AI Agents. This categorization allows us to analyze collective bot behaviors.  


\para{Ethical considerations.}  Data collection and analysis adhered to a study protocol approved by our local Institutional Review Board (IRB)\footnote{IRB information redacted for anonymous submission}. We followed all IRB guidelines for anonymizing and storing this dataset, including making significant efforts to exclude non-bot behavior from the dataset and preserve user privacy. We removed sessions originating from logged-in users from our institution or from institution-hosted networks, and anonymized IP addresses by hashing them. All data was stored and analyzed on secure servers only accessible by the research team.

\subsection{Dataset overview}

First, we present an overview of our dataset, examining both basic dataset statistics and bot behaviors of interest. Such an overview, independent of our work on \rbts compliance, is valuable. Little academic work has documented the contours of web scraper behavior in the AI age\textemdash and none have done so at scale. For our analysis, we aggregate dataset rows, each of which describes a single page access,  into time-based ``sessions'' associated with the same web agent. To do this, we identify accesses by a particular entity/user agent to a set of related pages at contiguous time steps and collapse them into one row, retaining information about individual subdomains visited in a session. We say a session ``ends" after 5 minutes of inactivity from an entity. This reduces our dataset from 3,914,096 rows to 761,956 rows. 

\para{High-level statistics.} Table~\ref{tab:sites_bots} gives an overview of our dataset, distinguishing between two subsets of data: data associated with any user agent (including generic ones), and data associated with known user agents with defined bot purposes. As the table shows, our dataset contains thousands of unique IP addresses and user agents, but only a handful are associated with known user agents. We define ``known bots'' as those with self-declared user agents that match either a user agent string format from our Github dataset using regular expression pattern matching, or one recorded by Dark Visitors. The low number of known bots relative to the total unique user agents in our dataset presages the difficulty of determining bot compliance with \rbts, since many operate anonymously or with generic user agents. 

\para{Most active bots.} Now, we focus specifically on known bots, e.g. bots with self-declared user agents that are well-documented by several web sources. Table~\ref{tab:top_botnames} gives details about the behavior of the 20 most active bots (as measured by number of web accesses) seen in our dataset. Together, these $20$ bots make up over $40\%$ of the observed web traffic. YisouSpider and AppleBot dominate, collectively driving $30\%$ of all traffic and scraping over $8$ GB of data during the 40 day period. YisouSpider is associated with the Chinese search engine Yisou and appears to be used to index web pages~\cite{yisouspider}, specifically focusing on educational and job related webpages ~\cite{darkvisitors}. We found that the vast majority of the YisouSpider's accesses were to our institution's people directory. AppleBot is a generic bot associated with Apple and is ``used by Apple to index search results that allow the Siri AI Assistant to answer user questions''~\cite{applebot}. 

We also investigate which categories of bots most frequently visit our sites, using the categories defined by Dark Visitors (see \S\ref{sec:data_prep}). Figure~\ref{fig:hist_bots} plots the relative frequency of different types of known bots in our dataset, separated into the categories defined by Dark Visitors. Search-related bots, either for traditional or AI-powered search, are the most frequent visitors, followed by AI data scrapers. Headless browsers, a category assigned to entities observed running a browser without a GUI, come in forth. This category is mostly composed of likely scraper bots that do not identify themselves with a known user agent string. Bots used for other applications, like SEO and archivers, are less common.

\para{Cumulative bot behaviors.} Finally, we explore how bots behave over time. Figure~\ref{fig:total_data} plots the CDF of data collected by the top $5$ categories of bots in terms of bytes scraped. From this, we see that most bots collect data at a steady rate, with search engine crawlers as a notable exception. These have a significant spike in data collection towards the end of our data collection period. Furthermore, we observe that AI assistants, though ranked sixth in of number of total scraping sessions (Figure~\ref{fig:hist_bots}) are ranked fourth in terms of data scraped (Figure~\ref{fig:total_data}), indicating that they collect a relatively large amount of data per session. 

Figure~\ref{fig:scraping_traffic_patterns} shows the number of sessions per day for the five categories of bots with the most total sessions. As with the data CDF plot, most categories of bots exhibit relatively stable behavior over time, with the exception of search engine crawlers and AI search crawlers. Upon further inspection,  we find these spikes in activity correspond directly with high scraping activity from YisouSpider (a search engine crawler) and AppleBot (an AI search crawler). 

\begin{figure*}
\begin{minipage}[t]{0.45\textwidth}
   \centering \input{figures/robots_baseline}
\end{minipage}\hfill
\begin{minipage}[t]{0.45\textwidth}
    \centering \input{figures/robots_v1}
\end{minipage}
\begin{minipage}[b]{0.48\textwidth}
    \centering \input{figures/robots_v2}
\end{minipage}\hfill
\begin{minipage}[b]{0.48\textwidth}
    \centering \input{figures/robots_v3}
\end{minipage}
\end{figure*}

\section{Measuring scraper compliance with \rbts}
\label{sec:active}

Now, we turn our attention to our central question: {\em understanding the extent to which scraper bots comply with \rbts directives}. We first describe our methodology, then analyze the results of our experiments. 


\subsection{Methodology} 
\label{sec:method}
Over a period of 8 weeks, we deployed four versions of \rbts files on one of the $36$ institution-controlled sites we were monitoring. Data from this site was also part of the dataset in \S\ref{sec:dataset}. This site was chosen because of its observed high bot traffic, making it a rich source of data on scraper behavior. Each \rbts file was deployed for two weeks, then swapped out by support staff for the next version. The four versions of the \rbts files have increasingly strict directives, ranging from no meaningful restrictions to full denial of access for certain bots. The restrictions were chosen to represent a steady gradient of \rbts behaviors, ranging from permissive to strict while only changing one condition at a time, allowing us to assess how bot behaviors change in response to incremental \rbts updates. We believe such controlled experimentation is the best approach, ensuring that confounding variables are not introduced by the modification of too many \rbts properties at once. Using these related but distinct versions of \rbts allowed us to achieve two distinct goals: measuring whether certain \rbts directives were more or less respected by scrapers, and measuring how quickly scrapers adapted to new \rbts restrictions. The files are described below and shown in Figures~\ref{fig:baseline_rbts}-\ref{fig:rbts3}.

 \begin{packed_itemize}
     \item {\bf Base version} (Fig.~\ref{fig:baseline_rbts}): All bots can access all but three pages (\texttt{/404, /dev-404-page, /secure/*}). This was our institution's standard \rbts setup.
    \item {\bf Version 1} (Fig.~\ref{fig:rbts1}): All bots retain the same level of access as in the base version, but a 30 second crawl delay is requested between successive page requests. 
    \item {\bf Version 2} (Fig.~\ref{fig:rbts2}): Most bots are allowed to access only the \texttt{/page-data/*} endpoint, which we observed experimentally to be a common target for scrapers, but are disallowed from all other sites. Eight search engine optimization (SEO) bots\footnote{Googlebot, Slurp, bingbot, Yandexbot, DuckDuckBot, BaiduSpider, DuckAssistBot, ia\_archiver} are exempt from these restrictions per our institution's request, to ensure the sites remain easily findable online. 
    \item {\bf Version 3} (Fig.~\ref{fig:rbts3}): Most bots are denied access to all page content. The same eight SEO bots as before are exempt. 
 \end{packed_itemize}

\noindent We validated that each \rbts file was formatted correctly via the Google \rbts parser\footnote{https://github.com/google/robotstxt}. Data collection associated with each of these files was conducted mostly in parallel with the data collection process described in the previous section. The one exception was the baseline \rbts data, which was collected in January 2025 before the full dataset collection started. Table~\ref{tab:rbts_stats} summarizes data collected under each \rbts version. Similar numbers of site visits and unique bots are observed across all four \rbts deployments.

 \begin{table}[h]
\centering
\begin{tabular}{lcc}
\toprule
\begin{tabular}[c]{@{}c@{}}\rbts\\ \bf version\end{tabular} &
  \begin{tabular}[c]{@{}c@{}}\bf unique site\\\bf visits\end{tabular} &
  \begin{tabular}[c]{@{}c@{}}\bf unique bot\\\bf visitors\end{tabular} \\ \midrule
     Base (Fig.~\ref{fig:baseline_rbts})   & $113,601$  &  $78$   \\
     v1 (Fig.~\ref{fig:rbts1}) &   $111,206$   &  $87$ \\
    v2 (Fig.~\ref{fig:rbts2}) & $119,399$ & $77$ \\
    v3 (Fig.~\ref{fig:rbts3}) &  $113,060$  &  $75$  \\ \bottomrule         
\end{tabular}%
\vspace{0.1cm}
\caption{{\bf Summary statistics of web traffic captured under all four \rbts versions.} Site traffic and number of unique bot visitors (bots with known user agents) remains consistent across all \rbts deployments.}
\label{tab:rbts_stats}
\vspace{-0.5cm}
\end{table}
 
\para{Data preparation.} We performed basic preprocessing on our datasets after collection to aid analysis. Much of our preprocessing pipeline for the \rbts experiments mirrored that described in \S\ref{sec:dataset}. However, for our analysis of compliance with \rbts we also filter out bots that accessed the site less than $5$ times under any \rbts version. Finally, we also eliminated any bots that appeared to have spoofed their user-agent. We determine potential spoofing by looking at bots for which $>90\%$ of traffic originates from a single ASN, and flag as possibly spoofed instances any requests originating from another ASN (e.g. not the $90\%$ majority ASN). A deeper exploration of this spoofing metric can be found in \S\ref{sec:mediating}.

\subsection{Metrics for \rbts compliance}
\label{sec:comply}

For each version of \rbts with some restrictions (v1, v2, v3), we develop novel metrics to determine if a bot complies with the directive(s), then use traditional statistical methods to measure whether such compliance is statistically significant relative to the bot's default behaviors. Here, we describe our compliance metrics. 

\para{Baseline \rbts.} We use web accesses associated with this version of \rbts as a control group to determine how/if bot behavior changes in response to stricter directives. 

\begin{table*}[htbp]
\centering
\begin{tabular}{llllc}
\toprule
\textbf{Bot category} & \textbf{Crawl delay} & \textbf{Endpoint access} & \textbf{Disallow all} & \textbf{Category average} \\
\midrule
AI Assistants           & 0.910 (2077) & 0.131 (2154) & {\bf 1.000} (1126) & 0.616 \\
AI Data Scrapers        & 0.560 (1857) & 0.352 (1824) & {\bf 0.766} (1818) & 0.559 \\
AI Search Crawlers      & {\bf 0.895} (1157) & 0.623 (974)  & 0.348 (1056) & 0.631 \\
Fetchers                & {\bf 0.925} (249)  & 0.283 (250)  & 0.377 (235) & 0.531 \\
Headless Browsers       & 0.036 (2777) & {\bf 0.278} (4207) & 0.011 (1282) & 0.155 \\
Intelligence Gatherers  & {\bf 0.809} (1120) & 0.372 (549)  & 0.094 (1008) & 0.450 \\
Other                   & {\bf 0.486} (5685) & 0.139 (3801) & 0.019 (3688) & 0.255 \\
SEO Crawlers            & 0.635 (1564) & {\bf 0.831} (1215) & 0.639 (1268) & {\bf 0.695} \\
Search Engine Crawlers  & {\bf 0.780} (6345) & 0.366 (5925) & 0.189 (6314) & 0.447 \\
\midrule
\textbf{Directive average} & \textbf{0.609} & 0.310 & 0.307 & 0.420 \\
\bottomrule
\end{tabular}
\vspace{0.1cm}
\caption{{\bf Bots are most likely to comply with the crawl delay directives, and SEO Crawlers are the most compliant bots overall.} We measure this by computing weighted averages of compliance ratios, weighted by number of bot accesses, for all bots in a particular category. We {\bf bold} the highest compliance value in each row to show directive-level trends in compliance and also {\bf bold} the highest row and column averages (last row/column).} 
\label{tab:weighted_avg}
\vspace{-0.3cm}
\end{table*}


\para{Crawl delay compliance (v1).} This \rbts version implements a 30 second crawl delay, the minimum time bots must wait between successive page accesses. To determine compliance with this directive, we first stratified our dataset into sets of accesses associated with a unique triple $\tau_i$ = (ASN, IP hash, user-agent), $i \in [0, K]$, where $K$ is the total number of $\tau$ tuples in our dataset.  This ensured we were correctly matching access logs with their unique requesting entity. We then sort each set of accesses by time and compute the time elapsed between each successive access.

We denote $\delta_{\tau_i} = \{\delta^j_{\tau_i} | j \in [0, n_i]\}$ as the set of access time deltas associated with tuple $\tau_i$, $n_i$ accesses total. Then, the {\em compliance ratio} for tuple $\tau_i$, $C_{\tau_i}$ is defined as
$$
C_{\tau_i} = \frac{|\{\delta^j_{\tau_i} | \delta^j_{\tau_i} \ge 30 \}|}{n}.
$$

\noindent In other words, we compute the ratio of access time deltas greater than or equal to $30$ seconds to total time deltas for tuple $\tau_i$. If $C_{\tau_i} = 1$, all access time deltas complied with the $30$ second crawl delay directive, and if $C_{\tau_i} = 0$, none did.  If there was only one access instance associated with a given $\tau_i$ triple, we count this as an instance of compliance, so $C_{\tau_i} = 1$.

After computing $C_{\tau_i}$ for the crawl-delay dataset, we run the same compliance ratio calculation on all access time deltas for $\tau$ triples in the baseline \rbts dataset, $C^{default}_{\tau_j}$, to determine the base rate of compliance with the $30$ second crawl delay directive. We then use a paired z-test for difference in proportions to determine if there is a statistically significant shift in the rate of compliance with the crawl delay directive for bots before/after the crawl-delay in \rbts is enforced. Since a single user agent (a term we use synonymously with bot) can be associated with multiple $tau_i$ tuples\textemdash for example, multiple IP addresses declare the same user-agent\textemdash we first group tuples based on user agent name, then analyze behavior across all $\tau_i$ tuples associated with this user agent. 


\para{Endpoint access compliance (v2).} This \rbts version restricts bots (except for a group of SEO bots, see \S\ref{sec:method}) to only accessing subpages under a single endpoint, \texttt{/page-data}, on the site. Computing the ratio of compliance with this directive is more straightforward. For each unique user-agent, we count its accesses to either the \rbts file (which is always allowed) or \texttt{/page-data} subpages and compute the ratio of this to the user agent's total page accesses. This compliance ratio $C_{\tau_i}$ is $1$ if the user-agent only accesses the allowed pages, \rbts and \texttt{page-data}, and is near $0$ if it consistently accesses many other pages. We then compute an identical compliance metric for user agents under our default \rbts and use this as the baseline for a paired z-test for differences in proportions between the two datasets. Ideally, we should see a significant uptick in compliance under the endpoint directive. 

\para{Disallow compliance (v3).} The final \rbts version prevents most bots (again, see exceptions in \S\ref{sec:method}) from accessing {\em any} endpoint on the page, although \rbts is always allowed. ``Compliance'' in this setting is therefore the ratio of \rbts accesses to total page accesses. If the bot behaves, all page accesses should be to \rbts, resulting in a compliance ratio of 1. Again, we compute this metric for both the dataset of disallow accesses and the baseline dataset and use a paired z-test for differences in proportion to measure change in compliance.

\subsection{Results and Discussion}
\label{sec:active_results}

In our analysis of \rbts compliance, we consider three main research questions: 

\begin{packed_itemize}
    \item {\bf RQ1:} With which directive are bots most likely to comply?
    \item {\bf RQ2:} Which category of bots have the highest rate of overall compliance?
    \item {\bf RQ3:} Do any individual bots exhibit interesting trends?
\end{packed_itemize}

To answer RQ1 and RQ2, we group bots by their Dark Visitors category and compute the {\em weighted} average of bot compliance with the $3$ directives for category. We weight the average by number of accesses from a particular bot, so the category average is more heavily weighted towards more common bots. We made this choice due to observed uneven distributions of bot accesses in our dataset. For example, some bots never comply with the directive but appear less than $10$ times, while others comply at a much higher rate and appear hundreds of times. A weighted average captures this nuanced behavior. Table~\ref{tab:weighted_avg} displays these results. 

We begin with RQ1. In Table~\ref{tab:weighted_avg}, we {\bf bold} the directive with the highest compliance rate for each bot. From this, we see that $5$ categories of bots comply the most with crawl delay, while $2$ categories of bots each have highest compliance with endpoint access and disallow all. Furthermore, we compute the average compliance rate for each directive, the last row in the table, and find the highest average compliance with crawl-delay\textemdash  nearly 2x that of the next-highest, endpoint access. Based on this analysis, we conclude that:

\begin{callout}
 {\bf Answer to RQ1:} Bots comply most with the crawl delay directive and least with the disallow all directive. This indicates that {\bf bots are less likely to comply with stricter \rbts directives}.
\end{callout}

\begin{figure*}
 \begin{minipage}{0.4\linewidth}
        \centering \includegraphics[width=1.0\textwidth]{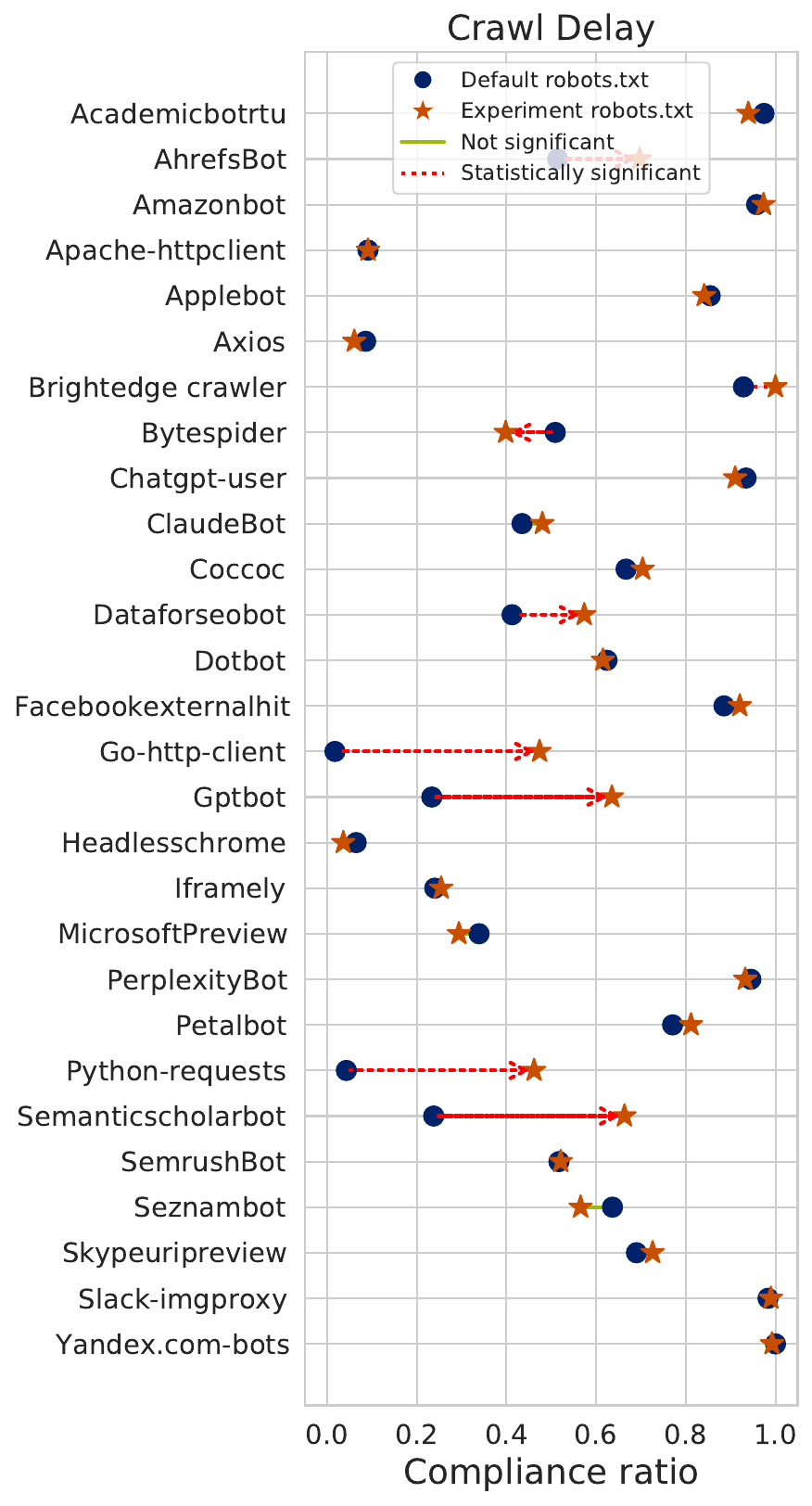}
    \end{minipage}\hfill
    \begin{minipage}{0.3\linewidth}
        \centering 
        \includegraphics[width=0.87\textwidth]{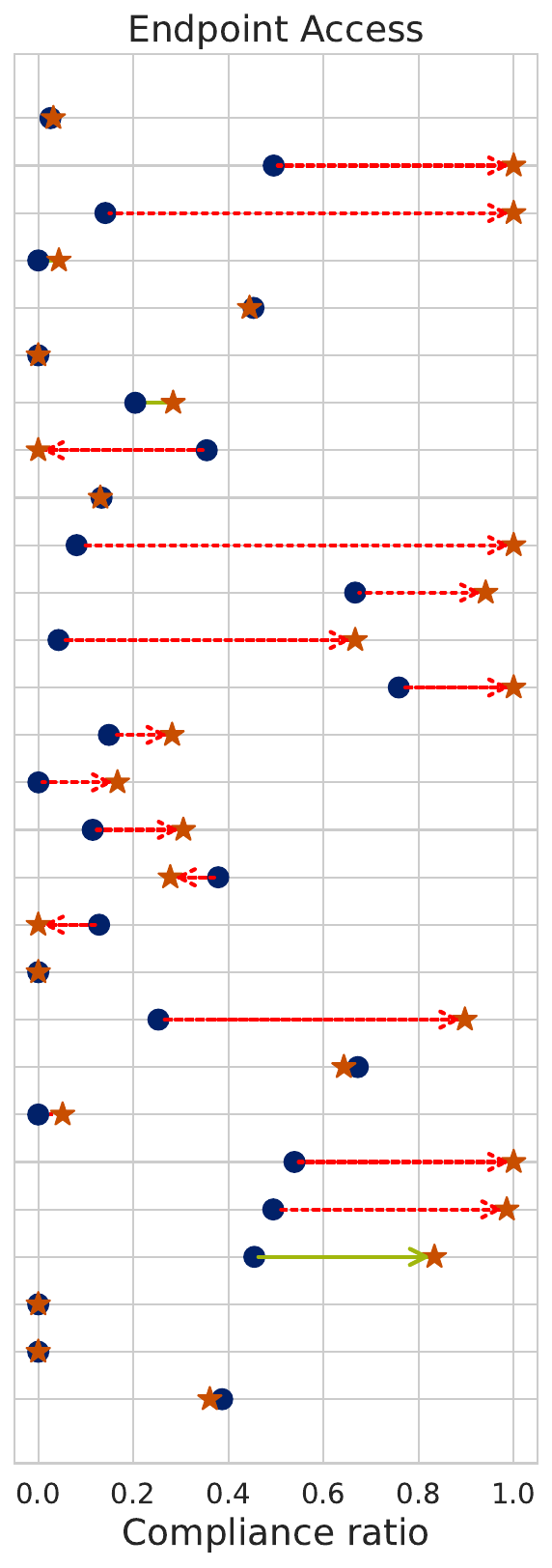}
    \end{minipage}
    \begin{minipage}{0.3\linewidth}
        \includegraphics[width=0.87\textwidth]{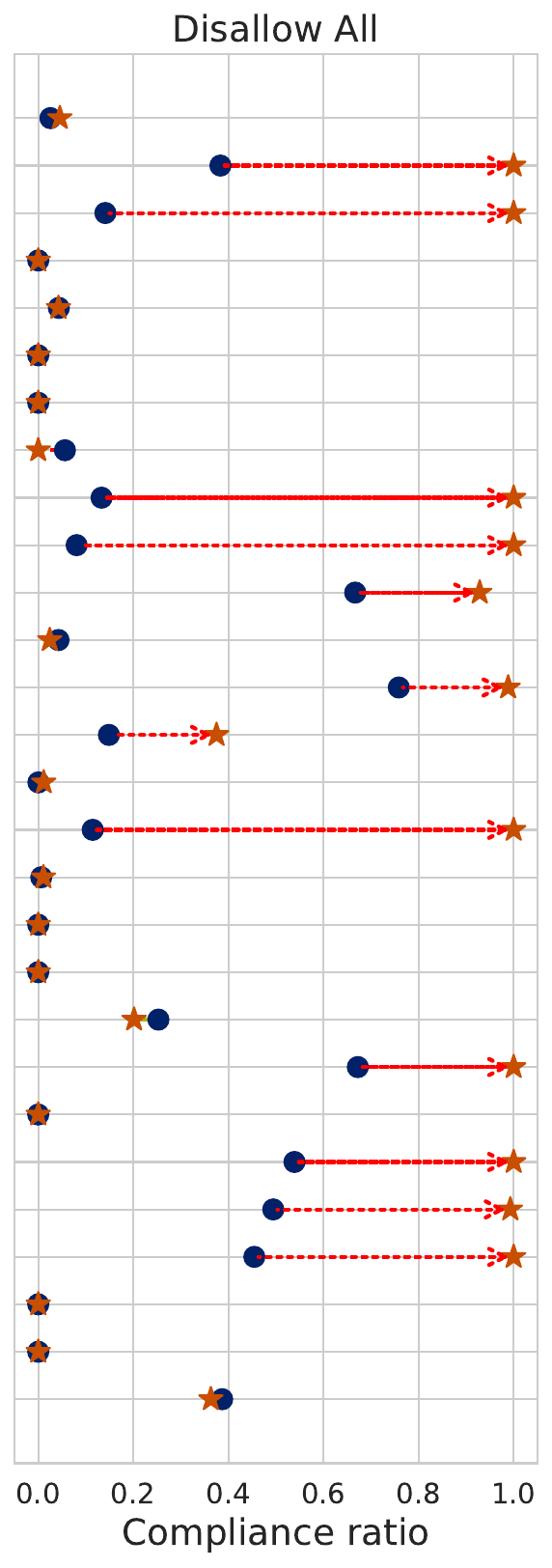}
    \end{minipage}
    \caption{{\bf Many bots comply by default with the 30 second crawl delay (leftmost graph), but the rate of compliance decreases overall as \rbts restrictions tighten.} We exclude results for exempted SEO bots and highlight statistically significant ($p \le 0.05$) shifts with red dotted lines. We also exclude results for bots that appear to have spoofed their user agent. Results for only putative spoofed bots are in Appendix~\ref{appx:spoofing}, and additional analysis of observed spoofing behavior is in  \S\ref{sec:mediating}.}
    \label{fig:compliance_ratios}
\end{figure*}

To answer RQ2, we compute the average compliance rate across all directives for each category and list this in the rightmost column of Table~\ref{tab:weighted_avg}. From this, we see that:

\begin{callout}
    {\bf Answer to RQ2:} SEO Crawlers exhibit the highest rate of compliance with \rbts directives, followed closely by AI Assistants and AI Search Crawlers.
\end{callout}

SEO Crawlers rely heavily on \rbts files to guide their actions, so their high compliance rate makes sense. The high compliance rates of AI Assistants and AI Search Crawlers are more surprisingly, especially given articles in the grey literature about AI bot {\em non-compliance} with \rbts~\cite{perplexity_wired, amazon-bots}. Further exploration of this disconnect is interesting future work. The least respectful bots, in contrast, are Headless Browsers and "Other." Since both these categories serve somewhat as a catchall, bots in this category are more obscure, making noncompliance unsurprising. 

\begin{table*}[htpb]
\centering
\resizebox{\textwidth}{!}{%
\begin{tabular}{lllccccc}
\toprule
\begin{tabular}[c]{@{}c@{}} \textbf{Bot Name} \end{tabular} & \begin{tabular}[c]{@{}c@{}} \textbf{Sponsoring} \\ \textbf{entity} \end{tabular} & \begin{tabular}[c]{@{}c@{}} \textbf{Bot} \\ \textbf{category} \end{tabular} & \begin{tabular}[c]{@{}c@{}} \textbf{Promise to} \\ \textbf{respect} \\ $\rbts$ \end{tabular} & 
\begin{tabular}[c]{@{}c@{}} \textbf{Crawl delay} \\ $\rbts$ \\ \textbf{compliance} \end{tabular} & \begin{tabular}[c]{@{}c@{}} \textbf{Endpoint} \\ $\rbts$ \\ \textbf{compliance} \end{tabular} & \begin{tabular}[c]{@{}c@{}} \textbf{Disallow} \\ $\rbts$ \\ \textbf{compliance} \end{tabular} \\ 
\midrule
Academicbotrtu                 & Riga Technical       & \colorbox{piedmont!30!white}{Other}      &  Unknown          & 0.939 & 0.032 & 0.045 \\
AhrefsBot                 & Ahrefs                     & \colorbox{whisper!95!black}{SEO}      & Yes           & 0.697 & 1.000 & 1.000 \\
Amazonbot                 & Amazon                     & \colorbox{dukeblue!30!white}{AI Search}    & Yes            & 0.973 & 1.000 & 1.000 \\
Apache-httpclient                 & Apache                     & \colorbox{piedmont!30!white}{Other}  &           Unknown    & 0.091 & 0.043 & 0.000 \\
Applebot                  & Apple                      & \colorbox{dukeblue!30!white}{AI Search}  & Yes              & 0.841 & 0.444 & 0.043 \\
Axios                & Open Source                     & \colorbox{piedmont!30!white}{Other}  & {\bf No}              & 0.060 & 0.000 & 0.000 \\
BrightEdge         & BrightEdge                 & \colorbox{whisper!95!black}{SEO}            & Yes           & 1.000 & 0.284 & 0.000 \\
Bytespider                & ByteDance                  & \colorbox{eno!40!white}{AI Data Scraper}       & {\bf No}          & 0.398 & 0.000 & 0.000 \\
ChatGPT-User              & OpenAI                     &  \colorbox{ironweed!40!white}{AI Assistant}          & Yes        & 0.910 & 0.131 & 1.000 \\
ClaudeBot                 & Anthropic                     & \colorbox{eno!40!white}{AI Data Scraper}  &           Yes    & 0.480 & 1.000 & 1.000 \\
Coccoc                    & Coc Coc                    & \colorbox{dandelion!50!white}{Search Engine}  & Yes          & 0.704 & 0.941 & 0.929 \\
DataForSEOBot             & DataForSEO                 & \colorbox{whisper!95!black}{SEO}            & Yes            & 0.573 & 0.667 & 0.024 \\
Dotbot                    & Moz                        & \colorbox{whisper!95!black}{SEO}            & Yes          & 0.615 & 1.000 & 0.988 \\
Facebookexternalhit       & Meta                       & \colorbox{copper!40!white}{Fetcher}  & {\bf No}             & 0.920 & 0.281 & 0.375 \\
Go-http-client       & Open Source                       & \colorbox{piedmont!30!white}{Other}  & Unknown             & 0.474 & 0.167 & 0.012 \\
GPTBot                    & OpenAI                     & \colorbox{eno!40!white}{AI Data Scraper}       & Yes           & 0.634 & 0.305 & 1.000 \\
Headlesschrome              & Open Source                       & \colorbox{shale!50!white}{Headless Browser}      &   Unknown         & 0.036 & 0.278 & 0.011 \\
Iframely              & Itteco                       & \colorbox{piedmont!30!white}{Other}      &   Yes        & 0.254 & 0.000 & 0.000 \\
MicrosoftPreview              & Microsoft                       & \colorbox{piedmont!30!white}{Other}      &    Yes        & 0.294 & 0.000 & 0.000 \\
PerplexityBot             & Perplexity                 & \colorbox{dukeblue!30!white}{AI Search}      & {\bf No}           & 0.933 & 0.897 & 0.202 \\
PetalBot                  & Huawei                     & \colorbox{dandelion!50!white}{Search Engine}  & Yes               & 0.812 & 0.643 & 1.000 \\
Python-requests              & Open Source                       & \colorbox{piedmont!30!white}{Other}      &       Unknown     & 0.462 & 0.051 & 0.000 \\
SemanticScholarBot        & Allen AI     & \colorbox{dandelion!50!white}{Search Engine}       & Yes            & 0.663 & 1.000 & 1.000 \\
SemrushBot                & Semrush                    & \colorbox{whisper!95!black}{SEO}            & Yes        & 0.521 & 0.986 & 0.993 \\
SeznamBot                 & Seznam.cz                  & \colorbox{dandelion!50!white}{Search Engine}  & Yes          & 0.565 & 0.833 & 1.000 \\
SkypeUriPreview           & Microsoft                  & \colorbox{piedmont!30!white}{Other}      & Yes              & 0.726 & 0.000 & 0.000 \\

Slack-imgproxy              & Salesforce                       & \colorbox{piedmont!30!white}{Other}      &     {\bf No}       & 0.917 & 0.000 & 0.000 \\

Yandex.com/bots           & Yandex                     & \colorbox{dandelion!50!white}{Search Engine}  & Yes         & 0.992 & 0.361 & 0.363 \\ \hline
\end{tabular}%
}
\caption{{\bf Individual bots respond differently to our \rbts directives.} We plot the top $26$ bots with $\ge 5$ accesses under each of our \rbts directives. For user agents associated with scraping libraries, we list ``open source'' as the sponsoring entity.}
\label{tab:individual_bot_compliance}
\vspace{-0.3cm}
\end{table*}

 Finally, we consider individual bot behavior in answering RQ3. Figure~\ref{fig:compliance_ratios} shows the change in compliance ratio from baseline for each directive across the $26$ bots with $\ge 5$ web accesses under each directive, denoting statistically significant shifts with a red dotted line. Table~\ref{tab:individual_bot_compliance} breaks out these bots by their category and includes information about their sponsoring organization and public promises made to respect \texttt{robots.txt}. From Figure~\ref{fig:compliance_ratios} we see a decreasing trend in baseline compliance with \rbts directives as we move from crawl delay to disallow all. This indicates that a number of bots complied by default with our 30 second crawl delay. Additionally, several bots make notable efforts to comply with each directive\textemdash Amazonbot, ClaudeBot, and GPTBot in particular. Finally, several bots (such as PerplexityBot) explicitly stated they will not respect \rbts (see Table~\ref{tab:individual_bot_compliance}) but have somewhat high compliance. This indicates either that they comply by default (which happened frequently for crawl delay) or are more respectful than they claim. There are also bots like BrightEdge that claim to respect \rbts but have low compliance.
\begin{callout}
{\bf Answer to RQ3}: There is significant variation in individual bot responses to \rbts directives. 
\end{callout}

\begin{table*}[th]
\centering
\resizebox{\textwidth}{!}{%
\begin{tabular}{ccccccc}
\toprule
\multirow{2}{*}{\textbf{Bot}} & \multicolumn{2}{c}{\bf Crawl delay \rbts} & \multicolumn{2}{c}{\bf Endpoint access \rbts} & \multicolumn{2}{c}{\bf Disallow all \rbts} \\ \cmidrule{2-7}
                              & Checked \rbts   & Compliance  & Checked \rbts    & Compliance    & Checked \rbts  & Compliance   \\ \midrule
Apache-HttpClient & No & 0.10 & Yes & 0.05 & No & 0.0 \\ 
Axios & No & 0.0 & No & 0.0 & No & 0.0 \\ 
BaiduSpider & No & 1.0 & No & 0.51$^*$ & No & 0.0$^*$ \\
BrightEdge Crawler    &       No       &    1.0         &      No               &        0.28       &    No               &     0.0 \\ 
Bytespider & Yes & 0.35 & No & 0.0 & Yes & 0.02 \\ 
DuckDuckBot & Yes & 0.07 & No & 0.0$^*$ & Yes & 0.02$^*$ \\
Googlebot-Image & No & 0.98 & No & 0.0$^*$ & No & 0.0$^*$ \\ 
Iframely & No & 0.17 & No & 0.0 & No & 0.0 \\ 
Microsoft-Preview & No & 0.13 & No & 0.0 & No &  0.0 \\
SkypeURIPreview & No & 0.67 & No & 0.01 & No & 0.0 \\ 
Slack-ImgProxy & No & 0.98 & No & 0.0 & No & 0.0 \\ \bottomrule

\end{tabular}%
}
\caption{{\bf Bots that skipped \rbts check during one or more \rbts experiments but still complied.} $^*$ = bot excluded from endpoint and disallow-all \rbts at institution's request.}
\label{tab:skipped_robots}
\vspace{-0.2cm}
\end{table*} 
\section{Exploring possible confounding variables in \rbts analysis}
\label{sec:mediating}



While the prior section shows that some bots respect \rbts, there are still several issues with relying on \rbts for restricting bot behavior. First, bots may simply not check \rbts frequently enough for changes in \rbts to be effective preventative measures. Second, bots may ``spoof'' their user agent, pretending to be a bot with high permissions in \rbts (e.g. Googlebot in our situation) to access otherwise-restricted content. In this section, we investigate how likely these two scenarios are in practice, to help practitioners understand practical limits of relying on \rbts to restrict bot behavior. 

\subsection{Frequency of \rbts updates}

\para{Motivation.}  There were several bots that simply never checked \rbts during our experiments: 9/34 for the crawl delay experiment, and 15/47 for both the endpoint access and the disallow-all experiments. Some of these bots still complied at least somewhat with the directives, but others did not comply at all. Table~\ref{tab:skipped_robots} lists all bots that did not check at least one of our experimental \rbts versions and their rates of compliance for each version. 

The case where bots never check \rbts but still comply (e.g. Slackbot, Googlebot-Image, Brightedge Crawler) raises some interesting questions. Do the bots comply without checking \rbts because (1) compliance happens naturally (e.g. directive is permissive enough) or (2) they retrieve \rbts from some other cached source? Without knowledge of crawler internal settings, both points are impossible to verify experimentally. We speculate that when bots comply with the crawl-delay directive without checking, compliance may have occurred naturally. It is unclear how compliance with stricter directives regarding page access could occur by chance, but investigation of the possibility of \rbts caching is out of scope for this paper. 
\begin{figure*}
    \centering
    \includegraphics[width=0.95\linewidth]{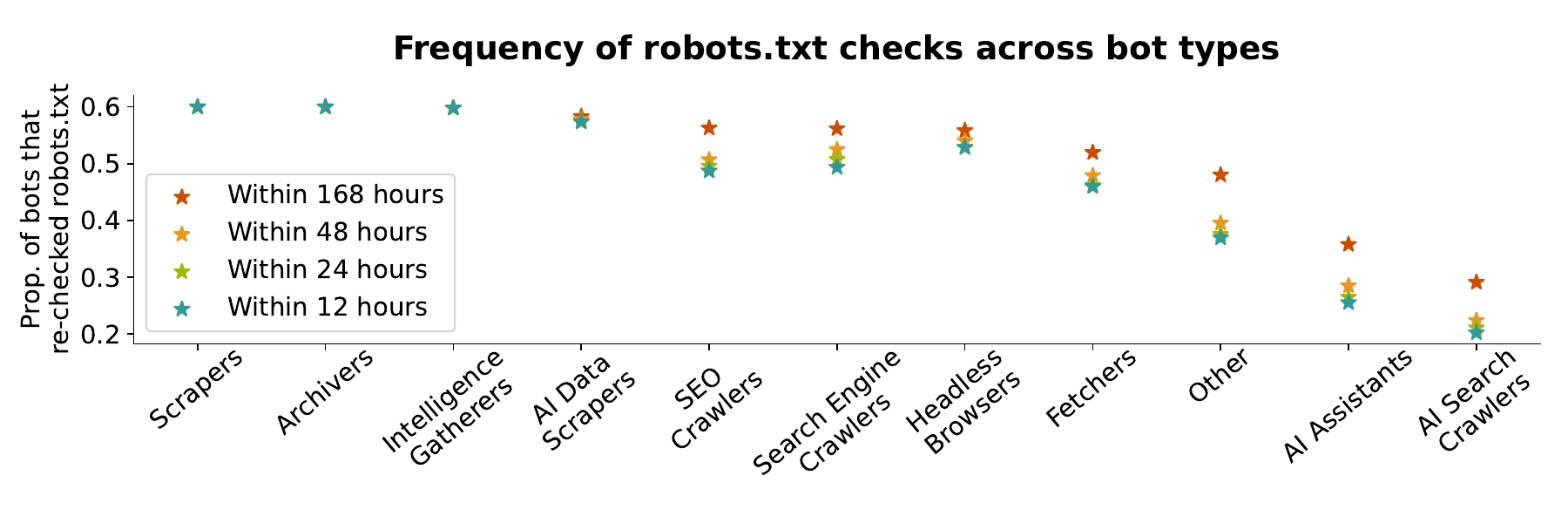}
    \vspace{-0.6cm}
    \caption{{\bf Some bots do not check \rbts very frequently.} AI assistants and AI search crawlers check \rbts the least.}
    \label{fig:24hr_cache_compliance}
    \vspace{-0.2cm}
\end{figure*}

The other interesting pattern in Table~\ref{tab:skipped_robots} is that some bots check \rbts under some conditions but not others (e.g. DuckDuckBot, Bytespider). This raises the question of whether some bots simply check \rbts infrequently. The standard set forth by Google (which other bots often follow) is that bots should check \rbts every 24 hours~\cite{google_rbts} but some bots may not follow this. Knowing which bots check \rbts less frequently on average could help web hosts determine how long they should expect to wait before bots adjust behavior based on their deployed \rbts and/or know if changing the \rbts file is the right approach for deterring a particular bot. 

\para{Methodology.} To understand the frequency with which bots check \rbts, we first find additional sites in our dataset for which \rbts were deployed with meaningful restrictions (e.g. contained at least one directive designed to change bot behavior) and were publicly available. Three out of the $36$ total sites had \rbts that met this criteria. All three files were identical and implemented simple restrictions on \texttt{/404} and $\texttt{/secure}$ endpoints on their respective sites. Through dataset inspection, we found that all bots respected these restrictions. 

To then analyze the frequency with which bots checked these files, we segmented the access logs for each bot into variable length time windows (12hrs, 24hrs, 48hrs, 72hrs, 168hrs) starting from when the bot first accessed any of these \rbts files. For each time window length, we then counted the number of windows in which the bot accessed \rbts throughout our 40 day dataset. We report a bot as ``complying'' with a particular time window access length if every time window of that length in the dataset contains a \rbts access. Using the variable length time windows allows us to see the bounds on \rbts frequency\textemdash do bots access it every $12$ hours? Every $24$? Every $168$?

We then grouped these results by bot category (e.g. Dark Visitors categories used previously), to give a better sense of aggregate bot behavior, and report the proportion of bots in each category that check \rbts within each of the $5$ time windows in Figure~\ref{fig:24hr_cache_compliance}.




\para{Results.} Some categories of bots (scrapers, archivers, and intelligence gatherers) consistently re-check \rbts within $12$  hours if they check it at all. Other categories exhibit more variable behavior, but increasing the window of time between \rbts checks typically leads to a higher re-check rate. AI-related bots (AI Assistants and AI search crawlers) have the lowest re-check rates with less than $40\%$ of bots checking \rbts within a $168$ hour window.  Thus, frequent updates to \rbts may not effectively prevent scraping, particularly for AI bots. 

\vspace{-0.2cm}
\subsection{User agent spoofing}
\label{subsec:spoof}

\para{Motivation.} Our analysis of \S\ref{sec:active} included checks for user agent ``spoofing.'' Here, we describe additional details of our methods for detecting spoofing and results. A key motivation for this investigation was our observation that a few bots had high-but-not perfect compliance rates. For example, Googlebot had a $0.65$ compliance rate with the 30 second crawl delay in \rbts, despite checking \rbts several times during this experiment. Similarly, Amazonbot had a $0.96$ compliance rate, even though it regularly checked \rbts. Several other popular bots like GPT-Bot and bingbot had similar trends.

This led us consider how much bot ``spoofing'' affected measured compliance rate for certain well-known bots. The useragent field in a web access request is manually set by the requesting agent, so malicious bots can easily manipulate this field to misrepresent themselves, potentially circumventing certain \rbts restrictions. 
 Some bots, like Googlebot, have elevated privileges in many \rbts files (including our experiments). By spoofing Googlebot's user agent, a bot could fly under the radar for longer before being banned. They could eventually be detected via IP address checks and behavioral patterns, but spoofing could help them avoid early flagging. 

Since a spoofing agent may not necessarily abide by the same rules as the benevolent bot, it may not, for example, respect \rbts. This could plausibly explain the partial non-compliance we saw with some of our \rbts directives. The grey literature gives some evidence of such spoofing in practice, particularly for large-scale AI bots~\cite{amazon-bots}.  

\para{Methods.} We analyzed web traffic data for known bots and observed empirically that most bots are overwhelmingly associated with a particular ASN. For example, $99.62\%$ of Googlebot's traffic comes from the \texttt{GOOGLE-CLOUD-PLATFORM} ASN, while $99.76\%$ of GPT-Bot's traffic is from \texttt{MICROSOFT-CORP}. Based on this, we develop an empirical heuristic that if a bot's traffic is associated $90\%$ of the time with one ASN, other ASNs associated with this user agent are likely spoofed. Using this heuristic, we analyze web accesses from our dataset and flag any bots associated with more than $1$ ASN for which one ASN has over $90\%$ of the traffic. Here, we focus specifically on well-known bots, for which spoofing is more interesting.  

\begin{table*}
\centering
\resizebox{\textwidth}{!}{%
\begin{tabular}{ccc}
\toprule
{\bf Bot} & {\bf Main ASN ($>90\%$ of accesses)} & {\bf Possible spoofing ASNs ($<5\%$ of accesses)} \\ \midrule
AdsBot-Google & GOOGLE & DMZHOST \\ \midrule
AhrefsBot & OVH & AHREFS-AS-AP \\ \midrule
Amazonbot & AMAZON-AES & CONTABO, DIGITALOCEAN-ASN \\ \midrule
Baiduspider & CHINA169-Backbone & \begin{tabular}[c]{@{}c@{}}CHINAMOBILE-CN, CHINANET-BACKBONE, CHINANET-IDC-BJ-AP,\\CHINATELECOM-JIANGSU-NANJING-IDC,\\ CHINATELECOM-ZHEJIANG-WENZHOU-IDC, HINET\end{tabular} \\ \midrule
bingbot & MICROSOFT-CORP-MSN-AS-BLOCK & \begin{tabular}[c]{@{}c@{}}Clouvider, HOL-GR,MICROSOFT-CORP-AS,\\ ORG-TNL2-AFRINIC, ORG-VNL1-AFRINIC\end{tabular} \\ \midrule
ClaudeBot & AMAZON-02 & GOOGLE-CLOUD-PLATFORM \\ \midrule
DuckDuckBot & MICROSOFT-CORP-MSN-AS-BLOCK & DIGITALOCEAN-ASN31, INTERQ31 \\ \midrule
facebookexternalhit & FACEBOOK &  \begin{tabular}[c]{@{}c@{}}AMAZON-02, AMAZON-AES,\\KAKAO-AS-KR-KR51\end{tabular} \\ \midrule
GPTBot & MICROSOFT-CORP-MSN-AS-BLOCK & BORUSANTELEKOM-AS \\ \midrule
Google Web Preview & GOOGLE & AMAZON-02 \\ \midrule
Googlebot-Image & GOOGLE & AMAZON-02 \\ \midrule
Googlebot/ & GOOGLE & \begin{tabular}[c]{@{}c@{}}52468, ASN-SATELLITE, ASN270353, CDNEXT,\\CHINANET-BACKBONE, Clouvider, DATACLUB,\\HOL-GR, HWCLOUDS-AS-AP, IT7NET, LIMESTONENETWORKS,\\M247, ORG-RTL1-AFRINIC, ORG-TNL2-AFRINIC,\\P4NET, PROSPERO-AS, RELIABLESITE,\\RELIANCEJIO-IN, ROSTELECOM-AS, ROUTERHOSTING,\\ TENCENT-NET-AP-CN, Telefonica\_de\_Espana, VCG-AS\end{tabular} \\ \midrule
meta-externalagent & FACEBOOK & DIGITALOCEAN-ASN \\ \midrule
SkypeUriPreview & MICROSOFT-CORP-MSN-AS-BLOCK & AMAZON-AES, M247\\ \midrule
Snap URL Preview Service & AMAZON-AES & AMAZON-02 \\ \midrule
Twitterbot & TWITTER & PROSPERO-AS, Telegram \\ \midrule

yandex.com/bots & YANDEX & \begin{tabular}[c]{@{}c@{}}AMAZON-02, AMAZON-AES,\\PROSPERO-AS\end{tabular} \\ \bottomrule
\end{tabular}
}
\caption{{\bf Popular bots with one dominant ASN and several infreqently-appearing ASNs\textemdash a possible sign of spoofing.}}
\label{tab:spoof}
\vspace{-0.3cm}
\end{table*}
\para{Results.} Overall, we identify $18$ bots for which spoofing may have occurred. In Table~\ref{tab:spoof}, we show the flagged bots, the dominant ASN (making up $\ge 90\%$ of bot traffic), and the suspicious ASNs ($\le 1\%$ of traffic). Although many flagged bots are only associated with two ASNs, some popular bots such as Googlebot are associated with up to $24$. Our analysis cannot prove that these infrequent ASNs belong to spoofers (maybe Google contracts with \texttt{Telefonica\_de\_Espana}?), but it strongly suggests this is the case. 

There are, on average, less than $5$ web accesses associated with these infrequent ASNs for most of the flagged bots. Notable exceptions are Baiduspider, with $381$ potentially spoofed accesses out of $15,132$; Googlebot, with $33$ out of $9103$; and SkypeURIPreview, with $26$ out of $831$. This means that spoofed bots may cause some, but not all, of the noncompliance with \rbts observed in \S\ref{sec:active}. We believe this relatively low spoofing rate could be due to tools deployed by our institution to filter out potential spoofers using whitelists of known IP addresses for common bots. Since these tools prevent certain suspicious bots from accessing institutional content a priori, we will not observe their traffic in our dataset. Nevertheless, the presence of potential spoofed bots that evade this analytical dragnet is interesting. 

\para{Limitations.} The study of spoofing presented here has several limitations. First, our threshold of $90\%$ of traffic originating from a single ASN is somewhat arbitrary. Future work could explore whether adjusting this threshold surfaces other interesting behaviors. Second, our analysis does not allow us to definitively state whether a bot is spoofing its user agent. Future work could consider alternative means, such as honeypots, that could enable more confident conclusions about which bot instances are spoofed.
\section{Discussion}
\label{sec:discussion}

From this large-scale study, we note several limitations of using \rbts as a deterrent for web scrapers, particularly AI scrapers. First, we observe that bots are less likely to comply with more restrictive \rbts directives, such as denying access to certain subpages, but are friendlier about respecting crawl delay. This implies that leveraging \rbts alone to prevent unwanted scraping could be ineffective. Second, we observed that many bots with known user agents in our data {\bf never checked the \rbts file} or {\bf check it infrequently}. This means that any changes made to \rbts to prevent unwanted scraping (e.g.~\cite{longpre2024consent}) would not be noticed by the scraper for significant time (if they are respected at all). Finally, we observe numerous instances of potential spoofing of well-known bots, indicating that malicious user-agents may attempt to evade \rbts restrictions by masquerading as bots with higher privileges. Together, these suggest that \rbts does not provide a universally respected signal for bot behavior, highlighting the need for more strongly-enforceable methods to prevent unwanted scraping. 

\para{Limitations of our study.} This study has several drawbacks that may limit the generalizability of its findings. The first study only considered traffic from $36$ websites owned by our institution. While these websites saw significant traffic during the course of data collection, they represent a small subset of sites online, and bots may behave differently on different types of sites (e.g. sites not affiliated with a large, well-resourced university). The second \rbts study was conducted on a single website, raising the possibility that bots behaved differently on this site than elsewhere. Both studies only capture a snapshot of bot behavior during a 45 day time window. Bot behavior will likely evolve over time, making it difficult to draw permanent lessons from our findings. Finally, in the first study, bot behavior was not static over time, which could impact findings from the second study.

\para{Future work.} Numerous threads of future work can build on this study. First, continued work is needed to find low-cost solutions to unwanted data scraping. This could take the form of a legally enforceable standard, a novel technical tool, or another new approach. Second, understanding needs of mid-size web hosts who might be exceptionally targeted by AI scraping (such as libraries and archives) through a user study could surface new, more targeted ways to provide more controlled access to this valuable data. Finally, additional work could dive into the spoofing analysis of \S\ref{sec:mediating}, determining more robust anti-spoofing methods that catch bots who slip through other enforcement mechanisms.

\newpage
\bibliographystyle{ACM-Reference-Format}
\balance
\bibliography{refs}

\newpage
\appendix

\section{Appendix}

\subsection{Ethics}
This paper raises few ethical concerns, besides those associated with potentially private data included in our user study. As specified in the paper, 
data collection and analysis adhered to a study protocol approved by our local Institutional Review Board (IRB). We followed all IRB guidelines for anonymizing and storing this dataset, including making significant efforts to exclude non-bot behavior from the dataset and preserve user privacy. We removed sessions originating from logged-in users from our institution or from institution-hosted WiFi networks, and anonymized IP addresses by hashing them. All data was stored and analyzed on secure servers only accessible by the research team.


\subsection{Spoofing analysis from \S\ref{sec:active_results}}
\label{appx:spoofing}

\begin{figure*}
 \begin{minipage}{0.33\linewidth}
        \centering \includegraphics[width=1.0\textwidth]{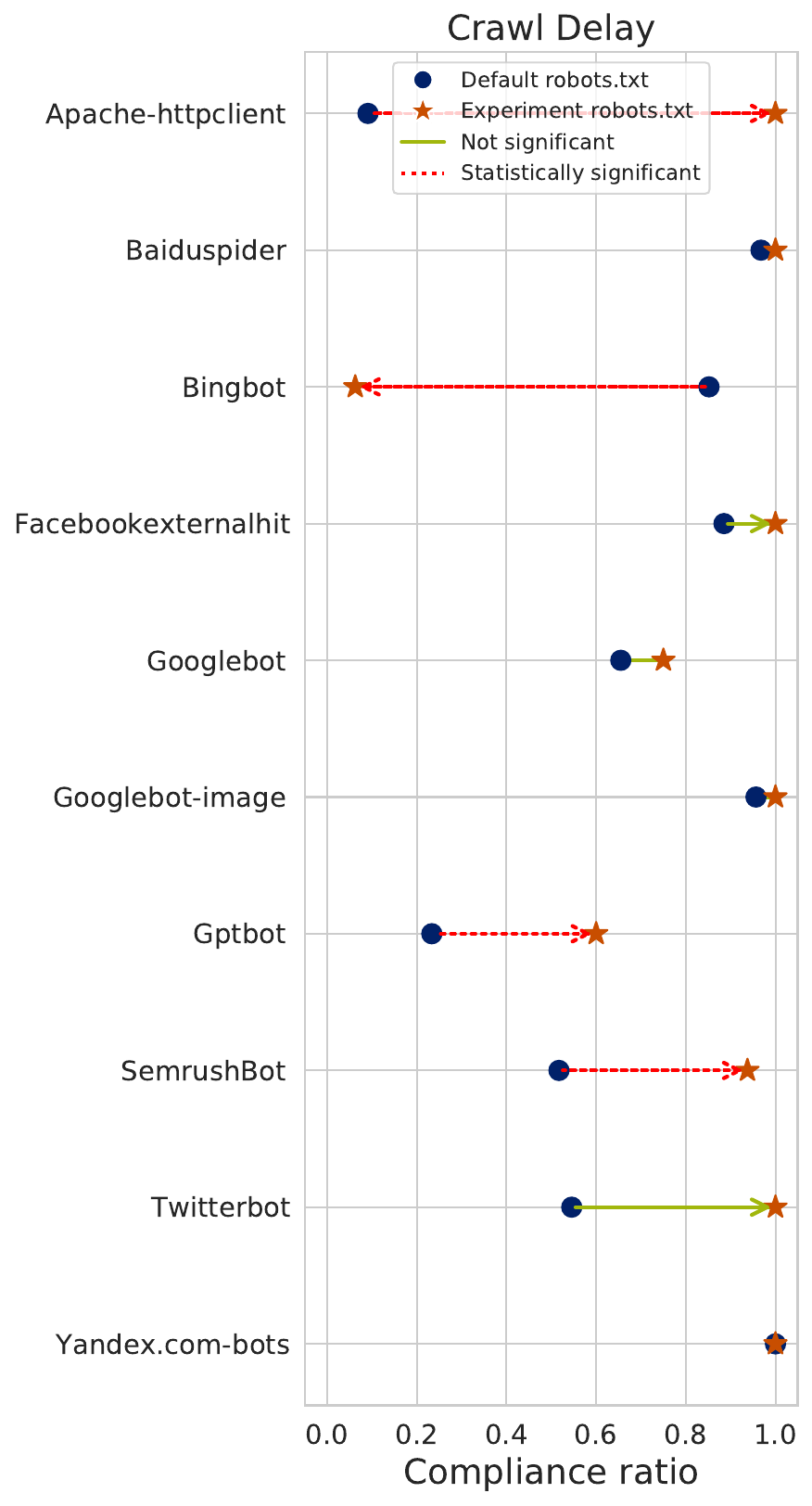}
    \end{minipage}\hfill
    \begin{minipage}{0.33\linewidth}
        \centering 
        \includegraphics[width=1.0\textwidth]{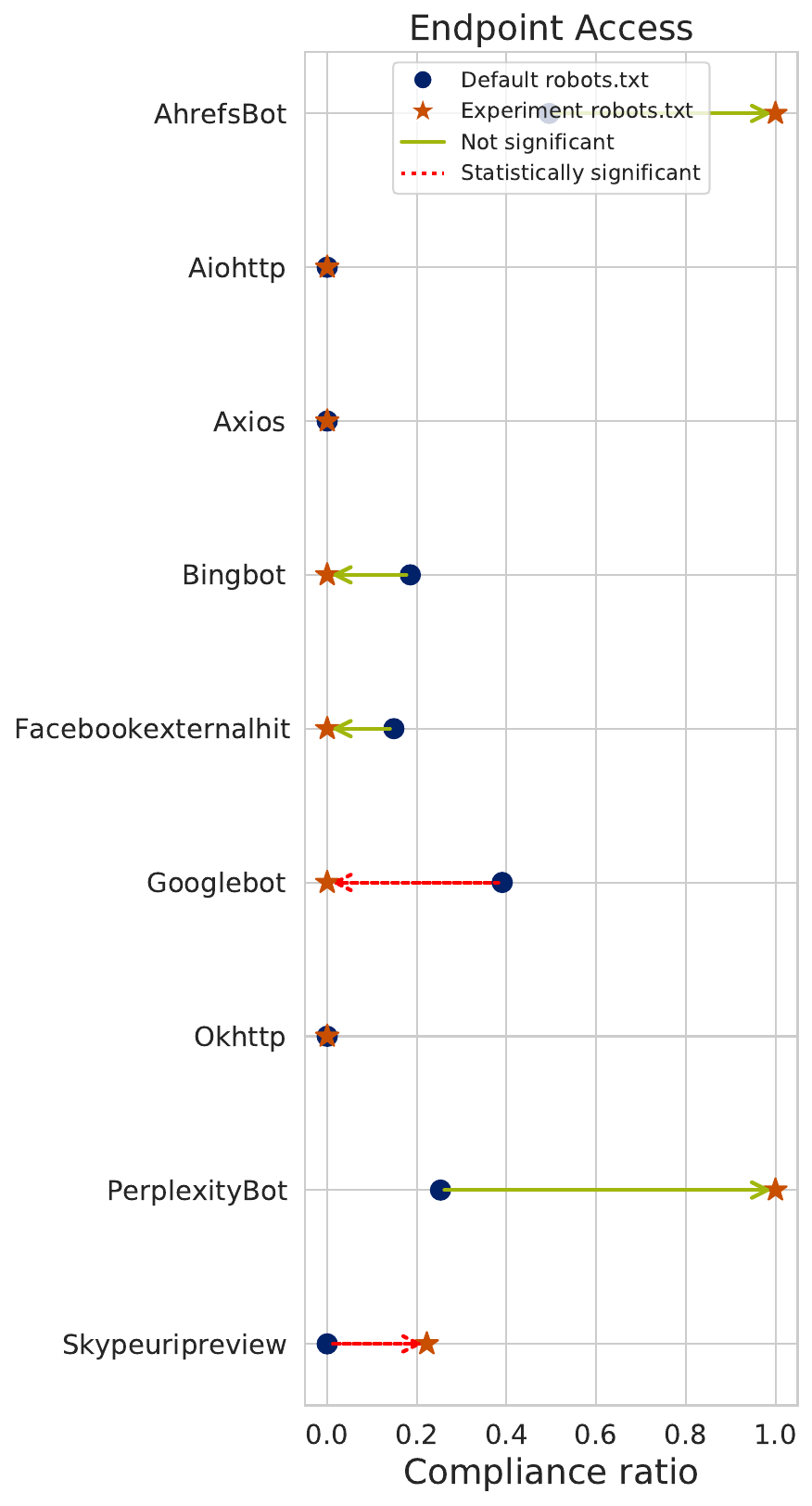}
    \end{minipage}
    \begin{minipage}{0.33\linewidth}
        \includegraphics[width=1.0\textwidth]{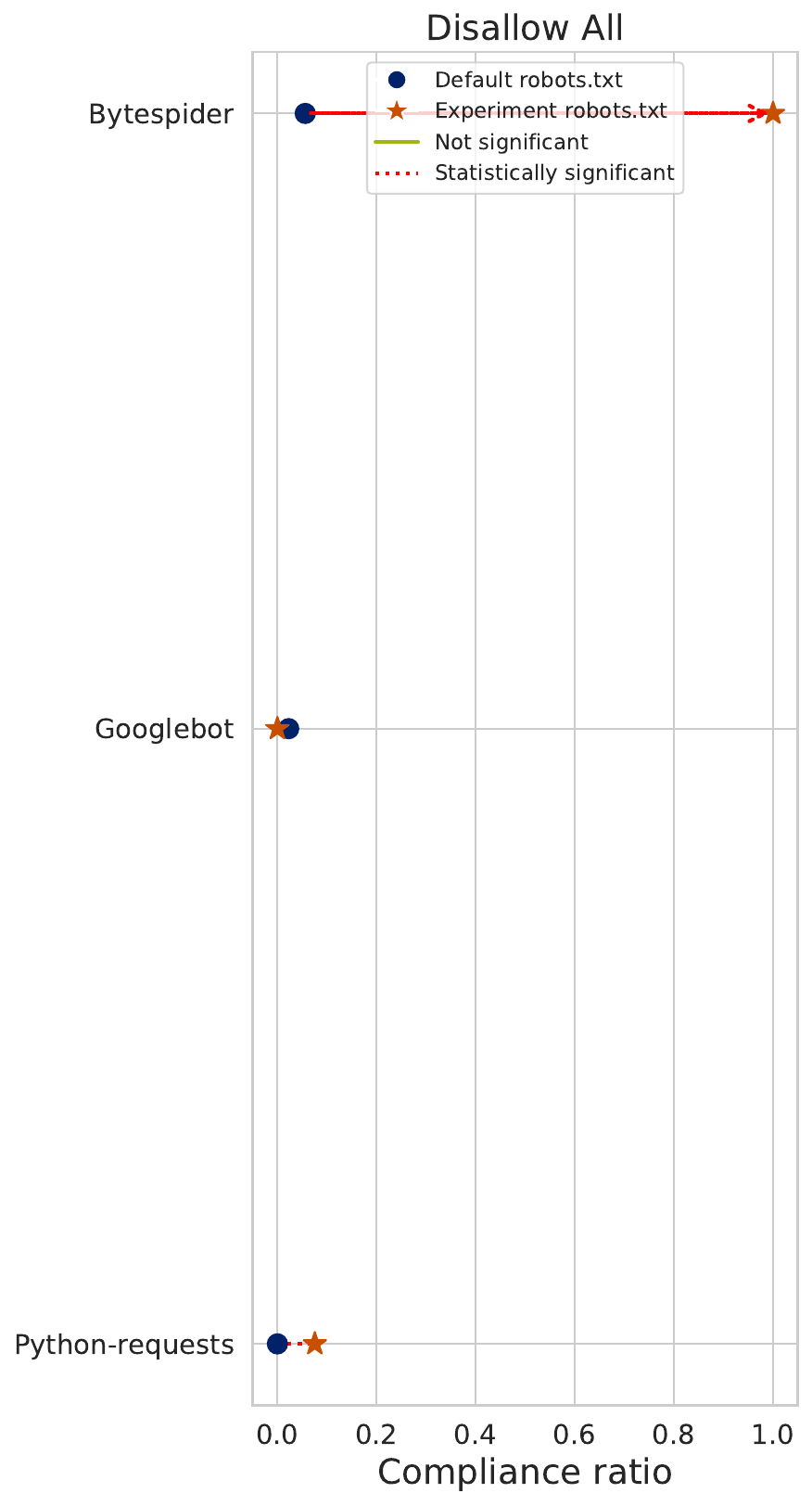}
    \end{minipage}
    \caption{{\bf The majority of potentially spoofed bots respond less\textemdash if at all\textemdash to \rbts~changes than their non-spoofed counterparts (see Figure~\ref{fig:compliance_ratios}).} The two exceptions are PerplexityBot and Bytedance bots under the Endpoint Access and Disallow All directives, which may be misidentified by our spoofing detection heuristic. We highlight statistically significant ($p \le 0.05$) shifts with red dotted lines.}
    \label{fig:spoofed_compliance_ratios}
\end{figure*}

Our \S\ref{sec:active_results} analysis of bot compliance with \rbts~excluded bots with apparently spoofed user agents. We used the methodology presented in \S\ref{sec:mediating} to identify potential spoofing bots: we look at bots for which $>90\%$ of traffic originates from a single ASN, and flag as possibly spoofed instances any requests with the same useragent originating from another ASN. In Figure~\ref{fig:spoofed_compliance_ratios}, we present parallel results to Figure~\ref{fig:compliance_ratios} but for putative spoofed bots. 

\begin{table*}[t]
    \centering
    \begin{tabular}{cccc}
    \toprule
       {\bf Directive}  &  {\bf Crawl delay} & {\bf Endpoint access} & {\bf Disallow all}  \\ \midrule
       Legitimate requests  & 10333 & 11397 & 8859\\ 
       Potentially spoofed requests & 60 & 36 & 247\\ 
       \bottomrule
    \end{tabular}
    \caption{{\bf Potentially spoofed requests account for less than $0.1\%$ of all traffic under our three crawl delay directives.} We report the number of perceived legitimate and potentially spoofed requests recorded during each of the three \rbts experiments.}
    \label{tab:spoof_counts}
\end{table*}

Overall, there were few instances of potentially spoofed requests flagged by our heuristic\textemdash see Table~\ref{tab:spoof_counts}. However, for the spoofed bots we did observe, there was little respect for \rbts~directives. Particularly for the endpoint access and disallow all directives, spoofed bot behaviors typically did not change before/after the directive was implemented. The two notable exceptions to this were PerplexityBot under the endpoint access directive and Bytespider under the disallow all directive. Both bots significantly changed their behavior, echoing the behavior seen in presumably true instances of the bots in Figure~\ref{fig:compliance_ratios}. This perhaps indicates that these were not spoofed instances, but rather instances of the true bot leveraging a different ASN than typically seen. Future work is needed to be able to definitively identify bots with spoofed user agents and avoid such ambiguity. 

\subsection{Statistical tests for change in bot compliance}

In Table~\ref{tab:zscores}, we report z-scores and p-values for the observed changes in compliance from the most common $26$ bots in our \rbts experiments. As is evident, many bots exhibit statistically significant changes in behavior in response to the various \rbts versions we deploy, indicating some level of awareness of the \rbts protocol and/or willingness to comply. 

\begin{table*}[htpb]
\centering
\begin{tabular}{lcccccc}
\toprule
\textbf{Bot Name} &
\multicolumn{2}{c}{\textbf{Crawl delay}} & 
\multicolumn{2}{c}{\textbf{Endpoint}} & 
\multicolumn{2}{c}{\textbf{Disallow}} \\
 & z-score & p-value & z-score & p-value & z-score & p-value \\ \midrule
AcademicBotRTU & -0.74 & 4.59e-01 & 0.18 & 8.59e-01 & 0.48 & 6.29e-01 \\
Amazonbot & 0.61 & 5.44e-01 & 10.71 & 0.00e+00 & 11.43 & 0.00e+00 \\
Apache-httpclient & 0.42 & 6.73e-01 & 0.99 & 3.21e-01 & N/A & N/A \\
Applebot & -0.45 & 6.49e-01 & -0.20 & 8.45e-01 & -0.04 & 9.66e-01 \\
Axios & -0.77 & 4.39e-01 & N/A & N/A & N/A & N/A \\
BrightEdge & 2.12 & 3.42e-02 & 1.21 & 2.25e-01 & N/A & N/A \\
Bytespider & -1.96 & 5.00e-02 & -5.04 & 4.74e-07 & -1.30 & 1.93e-01 \\
ChatGPT-User & -2.07 & 3.86e-02 & -0.18 & 8.60e-01 & 20.62 & 0.00e+00 \\
Claudebot & 0.44 & 6.62e-01 & 8.95 & 0.00e+00 & 8.24 & 2.22e-16 \\
Coccoc & 0.27 & 7.84e-01 & 2.67 & 7.51e-03 & 2.34 & 1.91e-02 \\
DataForSEOBot & 2.41 & 1.62e-02 & 6.35 & 2.14e-10 & -1.11 & 2.68e-01 \\
Dotbot & -0.17 & 8.63e-01 & 4.28 & 1.85e-05 & 4.66 & 3.12e-06 \\
Facebookexternalhit & 0.95 & 3.43e-01 & 2.49 & 1.27e-02 & 3.98 & 6.97e-05 \\
Go-http-client & 14.05 & 0.00e+00 & 19.24 & 0.00e+00 & 5.23 & 1.72e-07 \\
GPTBot & 11.54 & 0.00e+00 & 7.43 & 1.09e-13 & 24.20 & 0.00e+00 \\
HeadlessChrome & -3.15 & 1.63e-03 & -5.15 & 2.67e-07 & 0.87 & 3.83e-01 \\
Iframely & 0.26 & 7.97e-01 & -2.49 & 1.26e-02 & N/A & N/A \\
MicrosoftPreview & -1.00 & 3.15e-01 & N/A & N/A & N/A & N/A \\
PerplexityBot & -0.36 & 7.20e-01 & 6.86 & 7.01e-12 & -0.90 & 3.69e-01 \\
PetalBot & 0.58 & 5.64e-01 & -0.27 & 7.86e-01 & 2.81 & 4.94e-03 \\
Python-requests & 11.46 & 0.00e+00 & 5.04 & 4.64e-07 & 1.83 & 6.71e-02 \\
SemanticScholarBot & 10.39 & 0.00e+00 & 6.86 & 6.99e-12 & 7.36 & 1.81e-13 \\
SemrushBot & 0.91 & 3.62e-01 & 9.90 & 0.00e+00 & 10.18 & 0.00e+00 \\
SeznamBot & -0.49 & 6.26e-01 & 1.65 & 9.91e-02 & 2.83 & 4.65e-03 \\
SkypeUriPreview & 0.95 & 3.44e-01 & 1.66 & 9.61e-02 & N/A & N/A \\
Slack-imgProxy & 0.35 & 7.24e-01 & N/A & N/A & N/A & N/A \\
Yandex.com/bots & -1.77 & 7.64e-02 & -0.81 & 4.21e-01 & -0.74 & 4.58e-01 \\
\bottomrule
\end{tabular}
\caption{{\bf Statistical significance  of changes in compliance in response to each \rbts version.} We report z-scores and p-values for the $26$ most common bots across all experiments.}
\label{tab:zscores}
\vspace{-0.2cm}
\end{table*}

\end{document}